\documentclass[12pt,a4paper]{article}
\usepackage{a4wide}
\usepackage{amsmath}
\usepackage{graphicx}

\newlength{\absize}
\usepackage{verbatim}
\usepackage{cite}
\def\lsim{\mathrel{\raise.3ex\hbox{$<$\kern-.75em\lower1ex\hbox{$\sim$}}}}
\def\gsim{\mathrel{\raise.3ex\hbox{$>$\kern-.75em\lower1ex\hbox{$\sim$}}}}
\newcommand{\hc}{\hbox {h.c.}}
\usepackage[usenames]{color}

\catcode`@=11
\def\citer{\@ifnextchar [{\@tempswatrue\@citexr}{\@tempswafalse\@citexr[]}}

\def\@citexr[#1]#2{\if@filesw\immediate
  \write\@auxout{\string\citation{#2}}\fi
  \def\@citea{}\@cite{\@for\@citeb:=#2\do
    {\@citea\def\@citea{--\penalty\@m}\@ifundefined
       {b@\@citeb}{{\bf ?}\@warning
       {Citation `\@citeb' on page \thepage \space undefined}}%
\hbox{\csname b@\@citeb\endcsname}}}{#1}}
\catcode`@=12

\newcommand{\half}{{\textstyle\frac{1}{2}}}

\allowdisplaybreaks 

\begin{document}

\thispagestyle{empty}
\renewcommand{\thefootnote}{\fnsymbol{footnote}}
\newpage\normalsize
\pagestyle{plain}
\setlength{\baselineskip}{4ex}\par
\setcounter{footnote}{0}
\renewcommand{\thefootnote}{\arabic{footnote}}
\newcommand{\preprint}[1]{%
\begin{flushright}
\setlength{\baselineskip}{3ex} #1
\end{flushright}}
\renewcommand{\title}[1]{%
\begin{center}
\LARGE #1
\end{center}\par}
\renewcommand{\author}[1]{%
\vspace{2ex}
{\Large
\begin{center}
 \setlength{\baselineskip}{3ex} #1 \par
\end{center}}}
\renewcommand{\thanks}[1]{\footnote{#1}}
\renewcommand{\abstract}[1]{%
\vspace{2ex}
\normalsize
\begin{center}
\centerline{\bf Abstract}\par
\vspace{2ex}
\parbox{\absize}{#1\setlength{\baselineskip}{2.5ex}\par}
\end{center}}

\preprint{CERN-TH-2016-013}

\vspace*{4mm} 

\title{Spontaneous symmetry breaking in \\
the $S_3$-symmetric scalar sector} \vfill

\author{D. Emmanuel-Costa,$^{a,}$\footnote{E-mail: david.costa@tecnico.ulisboa.pt} 
O. M. Ogreid,$^{b,}$\footnote{E-mail: omo@hib.no}
P. Osland,$^{c,}$\footnote{E-mail: Per.Osland@ift.uib.no}  \\
M. N. Rebelo,$^{a,d,}$\footnote{E-mail: rebelo@tecnico.ulisboa.pt}}

\begin{center}
$^{a}$Centro de F\'isica Te\'orica de Part\'iculas -- CFTP\\
Instituto Superior T\'ecnico -- IST, Universidade de Lisboa, Av. Rovisco Pais, \\
P-1049-001 Lisboa, Portugal, \\
$^{b}$Bergen University College, Bergen, Norway, \\
$^{c}$Department of Physics and Technology, University of Bergen, \\
Postboks 7803, N-5020  Bergen, Norway \\
$^{d}$Theory Department, CERN, CH 1211 Geneva 23, Switzerland
\end{center}

\vfill

\abstract{ 
  We present a detailed study of the vacua of the $S_3$-symmetric
  three-Higgs-doublet potential, specifying the region of parameters
  where these minimisation solutions occur. We work with a CP
  conserving scalar potential and analyse the possible real and
  complex vacua with emphasis on the cases in which the CP symmetry
  can be spontaneously broken. Results are presented both in the
  reducible-representation framework of Derman, and in the
  irreducible-representation framework. Mappings between these are
  given.  Some of these implementations can in principle accommodate
  dark matter and for that purpose it is important to identify the
  residual symmetries of the potential after spontaneous symmetry
  breakdown.
We are also concerned with constraints from vacuum stability.
}

\vspace*{20mm} \setcounter{footnote}{0} \vfill

\newpage
\setcounter{footnote}{0}
\renewcommand{\thefootnote}{\arabic{footnote}}

\section{Introduction}
A possible direction in which to look for new physics beyond the
standard model emerges by enlarging the scalar sector, e.g., by adding
one or more scalar doublets. Models with two Higgs doublets
\cite{Gunion:1989we} have received a lot of attention (for a recent
review, see \cite{Branco:2011iw}). Attractive features of such
extensions are additional sources of CP violation \cite{Lee:1973iz,
Branco:1980sz, Branco:1985pf, 
Lavoura:1994fv, Botella:1994cs, Branco:1999fs, 
Branco:2005em, Gunion:2005ja, Nishi:2006tg,Accomando:2006ga,
Maniatis:2007vn, Grzadkowski:2013rza}, and a way to accommodate dark matter
\cite{Deshpande:1977rw,Barbieri:2006dq}.

Scalar sectors with three doublets have been considered, sometimes inspired by the existence of three generations of fermions. The general case involves a large number of free parameters
 \cite{Olaussen:2010aq}, and these parameters are
only constrained by general principles, like positivity, and a viable spectrum. Several authors, starting in 1977 with Pakvasa and Sugawara \cite{Pakvasa:1977in}, have imposed an $S_3$ permutation symmetry, in part motivated by a desire to model the fermion generations. As compared to the most general three-Higgs-doublet model, it is also attractive since it reduces the number of parameters.

Pakvasa and Sugawara exploited the reduction of $S_3$ to its
irreducible doublet and singlet. However, their potential, which has seven
quartic terms, was later \cite{Kubo:2004ps}
shown not to be the most general one, the term here referred to as the $\lambda_4$-term, was missing. 
In the sequel we show that this term plays a very important r\^ole.
An alternative formulation in terms of the three Higgs doublets in the reducible
representation transforming under the six three-by-three matrices of
permutation (to be referred to in the following as the reducible triplet) was presented by Derman \cite{Derman:1978rx} and further studied in a paper with Tsao \cite{Derman:1979nf}.

The descriptions of the potential in terms of the reducible-triplet and the irreducible frameworks are equivalent. But other sectors of the theory, in particular the Yukawa sector, would differentiate these frameworks and lead to different physics.

The vacua of the $S_3$-symmetric potential have been classified in terms of their residual symmetries by Ivanov and Nishi \cite{Ivanov:2014doa}. 
Here, we shall present another classification, namely in terms of constraints on the potential.
For real vacua, it was known that the condition $\lambda_4=0$ (see below) is relevant for the classification of the different vacua. We find that this parameter is also important for the complex vacua. Furthermore, there are several other constraints that are useful in this classification.
Such constraints are very important for model-building purposes.
Therefore,
we give a complete list of all possible vacua, starting from a scalar potential with real coefficients, and analyse their properties, giving the
constraints on the parameter space which allow for each solution.

The paper is organised as follows. In section~2 we set up some notation and discuss the potential. Section~3 is devoted to a general discussion of how the different vacua constrain the potential, whereas sections~4 and 5 present our results for the real and complex vacua. In section~6 we discuss connections between complex and real vacua, and in section~7 we briefly comment on the special case of $\lambda_4=0$, when the potential has an additional, continuous symmetry. Section~8 is devoted to a detailed discussion of spontaneous CP violation, and in section~9 we comment on dark-matter scenarios. Section~10 contains some concluding remarks. Technical issues are delegated to appendices A (mappings), B (positivity) and C (stationarity conditions).

\section{The $S_3$-symmetric potential}
\label{Sec:def-model}
\setcounter{equation}{0}

\subsection{Field notations}
We consider $S_3$-symmetric models with three $SU(2)\times U(1)$ reducible-triplet fields:
\begin{equation}
\phi_1, \quad \phi_2, \quad \phi_3.
\end{equation}
Allowing for complex vacuum expectation values, each field is decomposed as
\begin{equation} 
\phi_i=\left(
\begin{array}{c}\varphi_i^+\\ (\rho_i+\eta_i+i\chi_i)/\sqrt{2}
\end{array}\right), \quad i=1,2,3,
\label{Obasis}
\end{equation}
where $\rho_i$ is in general complex, whereas the fields $\eta_i$ and $\chi_i$ are real.

The reducible-triplet fields can alternatively be replaced by an $S_3$ doublet:
\begin{equation} \label{Eq:doublet}
\left(
\begin{array}{c}h_1\\ h_2
\end{array}
\right)
=\left(
\begin{array}{c}\frac{1}{\sqrt2}(\phi_1-\phi_2)\\ \frac{1}{\sqrt6}(\phi_1+\phi_2-2\phi_3)
\end{array}
\right),
\end{equation}
and an $S_3$ singlet 
\begin{equation}
h_S=\frac{1}{\sqrt3}(\phi_1+\phi_2+\phi_3),
\end{equation}
decomposed as
\begin{equation} \label{Eq:hi_hS}
h_i=\left(
\begin{array}{c}h_i^+\\ (w_i+\tilde \eta_i+i\tilde \chi_i)/\sqrt{2}
\end{array}\right), \quad i=1,2, \quad
h_S=\left(
\begin{array}{c}h_S^+\\ (w_S+\tilde \eta_S+i\tilde \chi_S)/\sqrt{2}
\end{array}\right),
\end{equation}
where again $w_i$ and $w_S$ can be complex.
Finally, the reducible-triplet fields may be replaced by a doublet and a pseudosinglet, denoted $h_A$, in which case the potential will take a slightly different form.

The potential, which has a quadratic and a quartic part,
\begin{equation}
V=V_2+V_4
\end{equation}
can be expressed either in terms of the reducible-triplet fields $\phi_i$, in terms of $h_1, h_2$, and $h_S$,  or in terms of $h_1, h_2$, and $h_A$. The first two formulations are equivalent.

It is useful to note that the (complex) vevs are related:
\begin{subequations} \label{Eq:DD-vacuum}
\begin{align}
w_1&=\frac{1}{\sqrt2}(\rho_1-\rho_2), \\
w_2&=\frac{1}{\sqrt6}(\rho_1+\rho_2-2\rho_3), \\
w_S&=\frac{1}{\sqrt3}(\rho_1+\rho_2+\rho_3),
\end{align}
\end{subequations}
with the inversion
\begin{subequations} \label{Eq:Derman-vacuum}
\begin{align}
\rho_1&=\frac{1}{\sqrt3}w_S+\frac{1}{\sqrt2}w_1+\frac{1}{\sqrt6}w_2, \\
\rho_2&=\frac{1}{\sqrt3}w_S-\frac{1}{\sqrt2}w_1+\frac{1}{\sqrt6}w_2, \\
\rho_3&=\frac{1}{\sqrt3}w_S-\frac{\sqrt2}{\sqrt3}w_2.
\end{align}
\end{subequations}

Whereas the formulation in terms of reducible-triplet fields is symmetric in $\phi_1, \phi_2, \phi_3$, 
the singlet--doublet representation is not. The decomposition into the doublet and singlet representations 
singles out a direction in terms of the $\phi$ fields. Any permutation of $\phi_i$ fields
in Eq.~(\ref{Eq:doublet}) would lead to an equally good definition for the components of the doublet.
This is a trivial fact. However,  this is the reason why  in the tables of possible vacuum states that follow, 
some cases that are equivalent in terms of  vacuum states of the reducible-triplet 
representation, given by (\ref{Eq:Derman-vacuum}),    have to be split into different cases in 
terms of those of the irreducible framework (\ref{Eq:DD-vacuum}) corresponding to different 
consistency conditions in terms of the minimisation of the potential.

\subsection{The potential in terms of reducible-triplet fields}
In terms of the reducible-triplet fields, the potential was written by Derman \cite{Derman:1978rx} as:
\begin{subequations}
\label{Eq:pot-original}
\begin{align}
V_2&=-\lambda\sum_{i}\phi_i^\dagger\phi_i +\half\gamma\sum_{i<j}[\phi_i^\dagger\phi_j+\hc],\\
V_4&=A\sum_{i}(\phi_i^\dagger\phi_i)^2
+\sum_{i<j}\{C(\phi_i^\dagger\phi_i)(\phi_j^\dagger\phi_j)
+\overline C (\phi_i^\dagger\phi_j)(\phi_j^\dagger\phi_i) 
+\half D[(\phi_i^\dagger\phi_j)^2+\hc]\} \nonumber \\
&+\half E_1\sum_{i\neq j}[(\phi_i^\dagger\phi_i)(\phi_i^\dagger\phi_j)+\hc]
+\sum_{i\neq j\neq k\neq i,j<k}
\{\half E_2[(\phi_i^\dagger\phi_j)(\phi_k^\dagger\phi_i)+\hc] 
\nonumber \\
&+\half E_3[(\phi_i^\dagger\phi_i)(\phi_k^\dagger\phi_j)+\hc] 
+\half E_4[(\phi_i^\dagger\phi_j)(\phi_i^\dagger\phi_k)+\hc]\}.
\end{align}
\label{29ab}
\end{subequations}
There are ten different coefficients in these equations. 

\subsection{The potential in terms of $S_3$ singlet and doublet fields}
\label{Sec:DasDey}
In terms of the $S_3$ singlet and doublet fields, the potential can be written as \cite{Kubo:2004ps,Teshima:2012cg,Das:2014fea}:
\begin{subequations} \label{Eq:V-DasDey}
\begin{align}
V_2&=\mu_0^2 h_S^\dagger h_S +\mu_1^2(h_1^\dagger h_1 + h_2^\dagger h_2), \\
V_4&=
\lambda_1(h_1^\dagger h_1 + h_2^\dagger h_2)^2 
+\lambda_2(h_1^\dagger h_2 - h_2^\dagger h_1)^2
+\lambda_3[(h_1^\dagger h_1 - h_2^\dagger h_2)^2+(h_1^\dagger h_2 + h_2^\dagger h_1)^2]
\nonumber \\
&+ \lambda_4[(h_S^\dagger h_1)(h_1^\dagger h_2+h_2^\dagger h_1)
+(h_S^\dagger h_2)(h_1^\dagger h_1-h_2^\dagger h_2)+\hc] 
+\lambda_5(h_S^\dagger h_S)(h_1^\dagger h_1 + h_2^\dagger h_2) \nonumber \\
&+\lambda_6[(h_S^\dagger h_1)(h_1^\dagger h_S)+(h_S^\dagger h_2)(h_2^\dagger h_S)] 
+\lambda_7[(h_S^\dagger h_1)(h_S^\dagger h_1) + (h_S^\dagger h_2)(h_S^\dagger h_2) +\hc]
\nonumber \\
&+\lambda_8(h_S^\dagger h_S)^2.
\label{Eq:V-DasDey-quartic}
\end{align}
\end{subequations}
(Teshima \cite{Teshima:2012cg} uses 
$(\lambda_1,\lambda_2,\lambda_3,\lambda_4,\lambda_5,\lambda_6,\lambda_7,\lambda_8)\leftrightarrow(C,D,G,E,B,F,F^\prime,A)$.)
In Appendix~A we give the translation between the parametrisations of the potential in terms of reducible-triplet fields and the one in terms of singlet and doublet fields.

Once again there are ten independent parameters. There are only four
terms in this potential that are sensitive to the relative phase of
different doublets, those in $\lambda_2$, $\lambda_3$, $\lambda_4$
and those in $\lambda_7$.  In terms of the reducible-triplet fields,
in Eqs.~(\ref{Eq:pot-original}) the number of such terms is higher
since here we have $\gamma$, $D$, $E_1$, $E_2$, $E_3$, and $E_4$.

In this formulation it is clear that the potential has an extra
$Z_2$ symmetry of the form $h_1 \rightarrow - h_1$. 
In terms of the equivalent doublet representation:
\begin{eqnarray}
\left( \begin{array}{c} \hat\chi_1 \\ \hat\chi_2 
\end{array} \right)
= \frac{1}{\sqrt{2}}\left( \begin{array}{cc}  
i  & 1  \\
-i & 1   
\end{array}\right)
\left( \begin{array}{c} h_1 \\ h_2
\end{array} \right),
\end{eqnarray}
which has also been used in the literature 
\cite{Chen:2004rr,Bhattacharyya:2010hp},
the above symmetry translates into a symmetry for the interchange
of the fields $\hat\chi_1$ and $\hat\chi_2$.

Another interesting feature is the fact that the choice  $\lambda_4=0$ leads to a 
continuous $SO(2)$ symmetry defined by:
\begin{eqnarray}
\left( \begin{array}{c} h_1^\prime \\ h_2^\prime
\end{array} \right)
= \left( \begin{array}{cc}
\cos \theta  & - \sin \theta  \\
\sin \theta & \cos \theta
\end{array}\right)
\left( \begin{array}{c} h_1 \\ h_2
\end{array} \right).
\label{so2}
\end{eqnarray}
This shows that the term with $\lambda_4$ plays a special r\^ole in the potential.
\subsection{The potential in terms of $S_3$ pseudosinglet and doublet fields}
\label{sect:pseudosinglet}
Instead of choosing the three Higgs doublets as being the singlet and the
doublet irreducible representations of $S_3$ we may choose them to be a pseudosinglet,
and the doubet.  These are also irreducible representations.
Under $S_3$ the pseudosinglet, $h_A$, transforms into $(- h_A)$. In this case there is
no direct translation of these fields into the defining reducible representation.

In terms of the $S_3$ pseudosinglet  and doublet fields, the potential
can be written as 
\begin{subequations}
\begin{align}
V_2&=\mu_0^2 h_A^\dagger h_A +\mu_1^2(h_1^\dagger h_1 + h_2^\dagger h_2), \\
V_4&=
\lambda_1(h_1^\dagger h_1 + h_2^\dagger h_2)^2 
+\lambda_2(h_1^\dagger h_2 - h_2^\dagger h_1)^2
+\lambda_3[(h_1^\dagger h_1 - h_2^\dagger h_2)^2+(h_1^\dagger h_2 + h_2^\dagger h_1)^2]
\nonumber \\
&+ \lambda_4[(h_A^\dagger h_2)(h_1^\dagger h_2+h_2^\dagger h_1)
-(h_A^\dagger h_1)(h_1^\dagger h_1-h_2^\dagger h_2)+\hc] 
+\lambda_5(h_A^\dagger h_A)(h_1^\dagger h_1 + h_2^\dagger h_2) \nonumber \\
&+\lambda_6[(h_A^\dagger h_1)(h_1^\dagger h_A)+(h_A^\dagger h_2)(h_2^\dagger h_A)] 
+\lambda_7[(h_A^\dagger h_1)(h_A^\dagger h_1) + (h_A^\dagger h_2)(h_A^\dagger h_2) +\hc]
\nonumber \\
&+\lambda_8(h_A^\dagger h_A)^2.
\label{Eq:V-David-quartic}
\end{align}
\end{subequations}

Apart from the ``trivial'' substitution $h_S\leftrightarrow h_A$, the two formulations (\ref{Eq:V-DasDey-quartic}) and (\ref{Eq:V-David-quartic}) differ in the $\lambda_4$-term, the two doublet fields are interchanged: $h_1\leftrightarrow h_2$.
Within the constraint of renormalizability (only quadratic and quartic terms) this scalar 
potential is equivalent to the previous one. However, this choice of representations will obviously have
implications for the Yukawa sector. We do not examine these implications in the
present work.

In the discussion of vacua, all results obtained for the irreducible framework in terms of the $S_3$ singlet and doublet can be trivially translated into this case. Therefore, our discussion will only refer to two different frameworks.
\subsection{Positivity}
Das and Dey have given {\it necessary} conditions for positivity \cite{Das:2014fea}. For the general potential, the {\it sufficient} conditions are rather involved. However, in the case of $\lambda_4=0$, they can be expressed quite explicitly, and are given in Appendix~B.

\section{The vacua---generalities}
\label{Sec:vacua-general}
\setcounter{equation}{0}

Since we are interested in CP violation, we will in general allow some vacuum expectation values (vevs) to be complex. However, due to the $U(1)$ invariance of the potential, one vev can always be chosen real. This holds in both frameworks.

The vacua can be determined from the conditions that derivatives of the potential with respect to the three independent fields must vanish.
These derivatives are linear in the coefficients of the potential, but cubic in terms of the (complex) vacuum expectation values. One approach would be to take the potential parameters as input, and solve these cubic equations for the vevs. 
In this section we shall follow another approach, which is to take the vevs as input, and use the derivatives to constrain the potential.
The quartic potential will also be constrained by positivity and an imposed particle spectrum.

We shall start this discussion by first quoting the minimisation conditions in the two frameworks. Clearly, one and the same vacuum will be phrased differently in the two frameworks. But one framework may give a simpler description than the other.

After writing out these derivatives in the next subsections, we shall first discuss how these conditions constrain the potential. Then (in section~\ref{Sec:RealVacua}), we review the real case (no CP violation), followed (in section~\ref{Sec:ComplexVacua}) by a discussion of the complex case, which may accommodate spontaneous CP violation.

\subsection{The reducible-triplet framework}

Within the reducible-triplet framework, three complex derivatives must vanish:
\begin{equation} \label{Eq:mincond}
\frac{\partial V}{\partial \rho_i^\ast}=0, \quad i=1,2,3,
\end{equation}
where
\begin{align}
\frac{\partial V}{\partial \rho_1^\ast}&=
\frac{-1}{2}\rho_1\lambda
+\frac{1}{4}(\rho_2+\rho_3)\gamma 
+\frac{1}{2}\rho_1^{\ast}\rho_1^2A
+\frac{1}{4}\rho_1(|\rho_2|^2+|\rho_3|^2)(C+\overline C) \nonumber \\
&+\frac{1}{4}\rho_1^\ast(\rho_2^2+\rho_3^2)D
+\frac{1}{8}[2|\rho_1|^2(\rho_2+\rho_3)
+\rho_1^2(\rho_2^\ast+\rho_3^\ast) 
+\rho_2^\ast\rho_2^2 + \rho_3^\ast\rho_3^2]E_1 \nonumber \\
&+\frac{1}{8}[\rho_1(\rho_2^\ast\rho_3 + \rho_2\rho_3^\ast)
+|\rho_2|^2\rho_3+\rho_2|\rho_3|^2](E_2+E_3) \nonumber \\
&+\frac{1}{8}(2\rho_1^\ast\rho_2\rho_3 
+ \rho_2^\ast\rho_3^2+ \rho_2^2\rho_3^\ast) E_4,
\end{align}
and $\partial V/\partial \rho_2^\ast$ and $\partial V/\partial \rho_3^\ast$ can be obtained by cyclic permutations.

We note that these derivatives do not depend on $C$ and $\overline C$ separately, only on the sum, $C+\overline C$. Likewise, they only depend on $E_2$ and $E_3$ via their sum. This means that the vacuum conditions are independent of the space spanned by the two parameters orthogonal to these, namely $C-\overline C$ and $E_2-E_3$. However, the spectrum will depend also on these parameters.
\subsection{The irreducible framework: singlet and doublet fields}
The three relevant derivatives that must vanish are now
\begin{align} \label{Eq:partial-hS}
\frac{\partial V}{\partial w_S^\ast}&=
\frac{1}{2}w_S\mu_0^2
+\frac{1}{4}[2|w_1|^2w_2+w_2^\ast(w_1^2-w_2^2)]\lambda_4 \nonumber \\
&+\frac{1}{4}w_S(|w_1|^2+|w_2|^2)(\lambda_5+\lambda_6)
+\frac{1}{2}w_S^\ast(w_1^2+w_2^2)\lambda_7
+\frac{1}{2}w_S^\ast w_S^2\lambda_8=0, \\
\label{Eq:partial-h1}
\frac{\partial V}{\partial w_1^\ast}&=\frac{1}{2}w_1\mu_1^2
+\frac{1}{2}w_1(|w_1|^2+|w_2|^2)\lambda_1
+\frac{1}{2}w_2(w_1^\ast w_2-w_1 w_2^\ast)\lambda_2 \nonumber \\
&+\frac{1}{2}w_1^\ast(w_1^2+w_2^2)\lambda_3
+\frac{1}{2}(w_1^\ast w_2 w_S+w_1 w_2^\ast w_S + w_1 w_2 w_S^\ast)\lambda_4
\nonumber \\
&+\frac{1}{4}w_1|w_S|^2(\lambda_5+\lambda_6)
+\frac{1}{2}w_1^\ast w_S^2\lambda_7=0, \\
\label{Eq:partial-h2}
\frac{\partial V}{\partial w_2^\ast}&=\frac{1}{2}w_2\mu_1^2
+\frac{1}{2}w_2(|w_1|^2+|w_2|^2)\lambda_1
-\frac{1}{2}w_1(w_1^\ast w_2-w_1 w_2^\ast)\lambda_2 \nonumber \\
&+\frac{1}{2}w_2^\ast(w_1^2+w_2^2)\lambda_3
+\frac{1}{4}[2(|w_1|^2-|w_2|^2)w_S+(w_1^2-w_2^2)w_S^\ast]\lambda_4
\nonumber \\
&+\frac{1}{4}w_2|w_S|^2(\lambda_5+\lambda_6)
+\frac{1}{2}w_2^\ast w_S^2\lambda_7=0.
\end{align}

We note that these derivatives do not depend on $\lambda_5$ and $\lambda_6$ separately, only on the sum, $\lambda_5+\lambda_6$. Likewise, they do not depend on $\lambda_1$, $\lambda_2$ and $\lambda_3$ separately, only on two combinations orthogonal to $\lambda_1+\lambda_2-2\lambda_3=0$.
\subsection{Constraining the potential by the vevs}
We are interested in the possibility of having spontaneous CP violation, therefore
we impose that  all the parameters of the potential should be real. Let us now consider 
the vevs as given a priori and solve the above minimisation conditions in terms
of parameters of the potential. Our basic discussion will be in the reducible triplet
framework. In this case the three vevs can be denoted as:
\begin{equation} \label{Eq:reducible-vev-notation}
\rho_i=v_ie^{i\tau_i},
\end{equation}
and we can write six minimisation conditions by computing the derivatives
of $V$ with respect to 
each of the $v_i$'s and of the $\tau_i$'s. It is clear from  Eqs.~(\ref{Eq:pot-original})
that, concerning phases, the potential  is only sensitive to  phase differences. In particular, we could 
choose without loss of generality a phase convention where one of these phases 
is rotated away, however, in this case we would loose symmetry among these equations. 
The explicit forms of these equations are given in Appendix~C.

As mentioned above, the pair of coefficients $C$ and
$\overline C$  as well as the pair $E_2$ and $E_3$ occur in each equation 
with a common factor, and therefore we are left with eight independent 
combinations of coefficients
and five independent real equations which should be chosen as the three equations
obtained from  $\partial V/\partial v_i = 0$ and any pair of those from
$\partial V/\partial \tau_i =0$.
We could in principle solve these equations for any set of five of the eight
independent parameters of the potential. These equations take the form:
\begin{eqnarray}  \label{five}
a_{11} P_1+a_{12} P_2 +a_{13} P_3 +a_{14} P_4 +a_{15} P_5 & = & b_1, \nonumber \\
a_{21} P_1+a_{22} P_2 +a_{23} P_3 +a_{24} P_4 +a_{25} P_5 & = & b_2, \nonumber \\
a_{31} P_1+a_{32} P_2 +a_{33} P_3 +a_{34} P_4 +a_{35} P_5 & = & b_3, \\
a_{41} P_1+a_{42} P_2 +a_{43} P_3 +a_{44} P_4 +a_{45} P_5 & = & b_4,  \nonumber \\
a_{51} P_1+a_{52} P_2 +a_{53} P_3 +a_{54} P_4 +a_{55} P_5 & = & b_5,  \nonumber 
\end{eqnarray}
where the $P_i$ denote different parameters of the potential. 
However not all of the possible $\binom{8}{5}=56$ combinations will lead to five independent 
equations. 

These five equations define five hyperplanes in the parameter space.
In the case of the reducible-representation framework, since $C$ and $\overline C$ appear together, as do $E_2$ and $E_3$, we have effectively an 8-dimensional parameter space.
Where the 5 hyperplanes intersect, we then have an $8-5=3$-dimensional parameter space, over which the vacuum is the same.

The requirement for the five equations to be independent
is that the determinant of the matrix $\cal A$
defined by:
\begin{equation}
{\cal A} 
= \left( \begin{array}{ccccc}
a_{11} & a_{12} & a_{13} & a_{14} & a_{15} \\
a_{21} & a_{22} & a_{23} & a_{24} & a_{25} \\
a_{31} & a_{32} & a_{33} & a_{34} & a_{35} \\
a_{41} & a_{42} & a_{43} & a_{44} & a_{45} \\
a_{51} & a_{52} & a_{53} & a_{54} & a_{55} 
\end{array} \right),
\label{calA}
\end{equation}
should be different from zero. It can readily be verified that the 
coefficients of the three
parameters $\lambda$, $A$ and $(C + \overline C)$ are not independent
and therefore these equations cannot be solved simultaneously
for these three parameters.  The terms with these coefficients are not 
sensitive to the relative phases and therefore they
do not appear in  the equations obtained from differentiating with respect to the phases. As a result, 
in order to check  this point it suffices to compute the $3\times3$ determinant 
involving the coefficients obtained from the first three minimisation conditions.
This determinant is zero. 

In the case of no spontaneous CP violation, the relative phases of the $\rho_i$ 
are zero and the corresponding minimisation condition,  
obtained from Eq.~(\ref{alphaa}) and cyclic permutations,
reduce to $0=0$ since each term in these equations is proportional 
to  the sine of relative phases. 
We are then left with only three independent equations and we can solve at most 
for three parameters of the potential.

Returning to the complex case,
we are now ready to classify the vacua, according to how many independent equations we have. 
In order for the five equations to be independent, it is sufficient that one of these 56 determinants be non-zero. Conversely, in order for at most four of the equations to be independent, all 56 possible such $5\times5$ determinants must vanish. 

For arbitrary vevs,
\begin{equation} \label{Eq:RRF-vacuum}
v_1, \quad
v_2e^{i\tau_2}, \quad
v_3e^{i\tau_3},
\end{equation}
we find that 16 out of the 56 possible $5\times5$ determinants vanish identically, whereas the remaining 40 are non-zero. The five equations (\ref{five}) can for any of these choices be solved in terms of the five parameters $P_1,\ldots,P_5$, with the exception of 5-parameter sets containing ($C,\overline C$), ($E_2,E_3$), ($\lambda, A,C$) or ($\lambda, \gamma, E_1, E_2$). The complements of these account for 14 out of the 16 vanishing ones. The remaining two are ($\gamma, A, C, E_1, E_2$) and ($\gamma, C, D, E_2, E_4$). In these sets, $C$ could be replaced by $\overline C$, and $E_2$ by $E_3$.

The remaining 40 determinants factorise, and vanish when either
\begin{gather}
\rho_i=0, \quad i=1,2,3 \quad \text{or} \\
\rho_i=\rho_j, \quad j\neq i.
\end{gather}
In these cases we can have at most 4 independent equations among the set (\ref{five}), and must investigate the corresponding $4\times4$ sub-determinants.

In the irreducible-representation framework, since $\lambda_5$ and $\lambda_6$ only appear as a sum in the minimisation conditions, we have effectively 9 parameters. Thus, we could have $\binom{9}{5}=126$ different $5\times5$ determinants. However, only 19 of these are non-vanishing. In this sense, this framework is more ``compact''. Here, the following parameter sets can not appear among the 5: ($\mu_0^2,\lambda_8$), ($\mu_1^2,\lambda_1$), ($\lambda_5,\lambda_6$), 
($\mu_0^2,\mu_1^2,\lambda_5$), ($\mu_0^2,\lambda_1,\lambda_5$), ($\mu_1^2,\lambda_2,\lambda_3$), ($\mu_1^2,\lambda_5,\lambda_8$), 
($\lambda_1,\lambda_2,\lambda_3$), ($\lambda_1,\lambda_5,\lambda_8$), 
($\mu_0^2,\lambda_2,\lambda_3,\lambda_5$), ($\lambda_2,\lambda_3,\lambda_5,\lambda_8$), and ($\lambda_3,\lambda_4,\lambda_5,\lambda_8$), as well as sets where in the above list $\lambda_5$ is replaced by $\lambda_6$. Among these, the sets ($\lambda_5,\lambda_6$), and ($\lambda_1,\lambda_2,\lambda_3$) correspond to ($C,\overline C$) and ($E_2,E_3$) in the reducible-representation framework. 

We shall distinguish the real and complex cases.

\section{Real vacua}
\label{Sec:RealVacua}
\setcounter{equation}{0}

For a real vacuum, the five equations (\ref{five}) discussed above reduce to a set of three.
Again, they are not necessarily all independent. If we, for example, try to solve for $\lambda$, $\gamma$ and $A$, the $3\times3$ determinant corresponding to (\ref{calA}) is particularly simple:
\begin{equation}
\det{\cal A}_{3\times3}
=-(\rho_1+\rho_2+\rho_3)(\rho_2-\rho_1)(\rho_3-\rho_2)(\rho_1-\rho_3).
\end{equation}
Thus, when this quantity is non-zero, we can solve for $\lambda$, $\gamma$ and $A$. Conversely, when $\det{\cal A}_{3\times3}=0$ (meaning the sum of the vevs is zero, or two are equal), then we have at most two independent equations, and can for example only solve for $\lambda$ and $\gamma$.

In the irreducible-representation framework, the three vacuum conditions (\ref{Eq:partial-hS})--(\ref{Eq:partial-h2}) can be solved to give $\mu_0^2$ and $\mu_1^2$ in
terms of the quartic coefficients:\footnote{There are misprints in the
  corresponding expressions given in Ref.~\cite{Das:2014fea}, their
  Eq.~(9): (i) a factor of 1/2 is missing on the right-hand side of
  all three expressions, and (ii) in $\mu_0^2$ ($\mu_3^2$ in their
  notation) the coefficient of $\lambda_4$ should be
  $(v_2/2v_3)(v_2^2-3v_1^2)$. These misprints were corrected 
in the Erratum provided by the authors and included in Ref.~\cite{Das:2014fea}.}
\begin{subequations} \label{Eq:DD-mu0-mu1}
\begin{align} \label{Eq:mu_0_sq}
\mu_0^2&=\frac{1}{2w_S}
\left[ \lambda_4(w_2^2-3w_1^2)w_2 -(\lambda_5+\lambda_6+2\lambda_7)(w_1^2+w_2^2)w_S -2\lambda_8w_S^3
\right], \\
\mu_1^2&=-\frac{1}{2}
\left[2(\lambda_1+\lambda_3) (w_1^2+w_2^2)+6\lambda_4w_2w_S +(\lambda_5+\lambda_6+2\lambda_7)w_S^2
\right], \label{Eq:DD-mu_1-a}\\
\mu_1^2&=-\frac{1}{2}
\left[2(\lambda_1+\lambda_3) (w_1^2+w_2^2)
-3\lambda_4(w_2^2-w_1^2)\frac{w_S}{w_2}
+(\lambda_5+\lambda_6+2\lambda_7)w_S^2 
\right]. \label{Eq:DD-mu_1-b}
\end{align}
\end{subequations}
The two equations (\ref{Eq:DD-mu_1-a}) and (\ref{Eq:DD-mu_1-b}) are not valid when $w_1 = 0 $ and $w_2 = 0$, since they were derived from (\ref{Eq:partial-h1})
and (\ref{Eq:partial-h2}) dividing by $w_1 = 0 $ and $w_2 = 0$, respectively. 
Furthermore, they are not automatically consistent. Consistency requires
\begin{subequations} \label{Eq:Das-Dey-condition}
\begin{align}
w_1&=0, \qquad \mbox{or else} \\
\lambda_4(3w_2^2-w_1^2)w_S&=0.
\label{Eq:lambda4=0,DD}
\end{align}
\end{subequations}

For $w_1=0$ the derivative of the potential with respect to $w_1$ is identically zero and therefore there is no clash in the determination of $\mu_1^2$ from the derivative with respect to $w_2$. From equation (\ref{Eq:lambda4=0,DD}) we see that these two derivatives are consistent if either $\lambda_4=0$ or $w_1=\pm\sqrt{3}w_2$ or else $w_S=0$.
The case $w_S=0$ is special since if we now take into account the derivative of the potential with respect to $w_S$, which is given by Eq.~(\ref{Eq:partial-hS}), we are left in the real case with
\begin{equation}
\lambda_4 w_2(3w_1^2-w_2^2)=0,
\label{4.4}
\end{equation}
which is the only term in Eq.~(\ref{Eq:partial-hS}) where $w_S$ does not appear as a factor.
As a result, solutions with $w_S=0$ require in addition that $\lambda_4=0$ or $w_2=\pm\sqrt{3}w_1$, or else $w_2=0$.
See cases R-I-2 in Table~\ref{Table:real}. These do not require $\lambda_4=0$, while case R-II-3 has $w_S=0$ and requires $\lambda_4=0$.

The different solutions can be summarised as given in Table~\ref{Table:real}, where the descriptions in terms of both the reducible- and irreducible-representation frameworks are given. For the purpose of making this table as well as the corresponding one for complex vacua more compact, we introduce the abbreviations
\begin{subequations} \label{Eq:lambda_ab}
\begin{align}
\lambda_a&=\lambda_5+\lambda_6+2\lambda_7, \\
\lambda_b&=\lambda_5+\lambda_6-2\lambda_7.
\end{align}
\end{subequations}

\begin{table}[htb]
\caption{Possible real vacua (partly after Derman and Tsao \cite{Derman:1979nf}).
The classification of vacua uses the notation R-X-y,
where R means that the vacuum is real. The roman numeral X is the number of constraints on the parameters of the potential that arise from solving the stationary-point equations. The letter y is used for distinguishing different vev's that have the same X, and $\lambda_a$ is defined in Eq.~(\ref{Eq:lambda_ab}).}
\label{Table:real}
\begin{center}
\begin{tabular}{|c|c|c|c|}
\hline
\hline
Vacuum  & $\rho_1,\rho_2,\rho_3$ & $w_1,w_2,w_S$ & Comment  \\
\hline
\hline
R-0& $0,0,0$  & $0,0,0$ & Not interesting  \\
\hline
\hline
R-I-1 & $x,x,x$  & $0,0,w_S$ & $\mu _0^2= -\lambda _8 w_S^2$  \\
\hline
R-I-2a & $x,-x,0$ & $w,0,0$ &  $\mu _1^2=- \left(\lambda _1+\lambda _3\right) w_1^2$\\
\hline
R-I-2b & $x,0,-x$ & $w, \sqrt{3} w, 0$ & $\mu _1^2= -\frac{4}{3} \left(\lambda _1+\lambda _3\right) w_2^2$ \\
\hline
R-I-2c & $0,x,-x$ & $w, -\sqrt{3} w, 0$ &$\mu _1^2= -\frac{4}{3}\left(\lambda _1+\lambda _3\right) w_2^2$  \\
\hline
\hline
R-II-1a & $x,x,y$ & $0,w,w_S$ & $\mu _0^2= \frac{1}{2}\lambda _4\frac{ w_2^3}{w_S}
-\frac{1}{2} \lambda_a w_2^2-\lambda _8 w_S^2$, \\
& & & $\mu _1^2= -\left( \lambda _1+ \lambda _3\right) w_2^2+\frac{3}{2} \lambda _4 w_2 w_S-\frac{1}{2} \lambda_a w_S^2$\\
\hline
R-II-1b & $x,y,x$ & $w,-w/\sqrt{3},w_S$ & $\mu _0^2= -4\lambda _4\frac{ w_2^3}{w_S}-2\lambda_a w_2^2-\lambda _8 w_S^2$, \\
& & & $\mu _1^2=-4 \left(\lambda _1+\lambda _3\right) w_2^2 -3 \lambda _4 w_2 w_S-\frac{1}{2} \lambda_a w_S^2$\\
\hline
R-II-1c & $y,x,x$ & $w,w/\sqrt{3},w_S$ &  $\mu _0^2= -4\lambda _4\frac{w_2^3}{w_S}-2\lambda_a w_2^2-\lambda _8 w_S^2$, \\
& & & $\mu _1^2= -4 \left(\lambda _1+\lambda _3\right) w_2^2-3 \lambda _4 w_2 w_S-\frac{1}{2} \lambda_a w_S^2$\\
\hline
R-II-2 & $x,x,-2x$ & $0, w, 0$ &$\mu _1^2= - \left(\lambda _1+\lambda _3\right) w_2^2$, $\lambda_4=0$  \\
\hline
R-II-3 & $x,y,-x-y$ & $w_1,w_2,0$ & $\mu _1^2= -\left(\lambda _1+\lambda _3\right)(w_1^2+w_2^2),\lambda_4=0$ \\
\hline
\hline
R-III & $\rho_1,\rho_2,\rho_3$ & $w_1,w_2,w_S$ & $\mu _0^2= -\frac{1}{2} \lambda_a( w_1^2+ w_2^2)-\lambda _8 w_S^2$,\\
& & & $\mu _1^2= - \left( \lambda _1+ \lambda _3\right)( w_1^2+ w_2^2)-\frac{1}{2} \lambda_a w_S^2$, \\
& & & $\lambda_4=0$ \\
\hline
\hline
\end{tabular}
\end{center}
\end{table}

One should note that
\begin{itemize}
\item
Vacuum R-I-1 is a special case of Vacuum R-II-1. In this case, the vacuum value $x$ is determined by
\begin{equation} \label{Eq:Vac-I-Derman}
\lambda-\gamma
=x^2[A+C+\overline C+D+2E_1+E_2+E_3+E_4].
\end{equation}
\item
For Vacuum~R-I-1, in the irreducible framework, we have
\begin{equation}
\mu_0^2=-w_S^2\lambda_8,
\end{equation}
which corresponds to Eq.~(\ref{Eq:Vac-I-Derman}), with $w_S^2=3x^2$.
\item
In the Vacua~R-I-2a, R-I-2b, R-I-2c, the vacuum value $x$ is determined by:
\begin{equation} \label{Eq:Vac-I-a-b-c-Derman}
2\lambda+\gamma=x^2[2A+C+\overline C+D-2E_1].
\end{equation}
In the irreducible framework, $\mu_0^2$ is not constrained by Eq.~(\ref{Eq:mu_0_sq}), whereas $\mu_1^2=-\lambda-\half\gamma$ is determined by the minimisation condition given above.

Vacua~R-I-2a, R-I-2b and R-I-2c, which correspond to $w_S=0$, require special discussion. It is clear from Eq.~(\ref{Eq:mu_0_sq}) that $\mu_0^2$ remains undetermined. According to Eqs.~(\ref{Eq:DD-mu_1-a}) and (\ref{Eq:DD-mu_1-b}), in these vacua the following relation must hold:
\begin{equation}
\mu_1^2=-(w_1^2+w_2^2)(\lambda_1+\lambda_3).
\end{equation}
(Special cases are given in Table~\ref{Table:real}.)
Using the translation given in Appendix~A, and taking $2x^2=w_1^2+w_2^2$, this constraint is seen to be equivalent to (\ref{Eq:Vac-I-a-b-c-Derman}).

\item
There are also solutions with $(\rho_1,\rho_2,\rho_3)=$ $(x,x,-2x),$ $(x,-2x,x),$ and $(-2x,x,x).$ These are reminiscent of vacua R-I-2a, R-I-2b and R-I-2c, with the interchange of $w_1$ and $w_2$.
\item
In the Vacua~R-II-1 and R-III, the two  coefficients of the bilinear potential, ($\mu_0^2$ and $\mu_1^2$) or ($\lambda$ and $\gamma$), can be determined from chosen vacuum values, together with the quartic potential. 
\item
Vacua~R-II are characterised by {\it two} independent vevs, referred to as $x$ and $y$ in the reducible-triplet framework, and as $w$ and $w_S$ in the irreducible framework. In the
framework of the reducible-triplet representation the three permutations presented as
subcases a), b) and c) are trivial, however, it should be pointed out that in the 
irreducible framework different consistency conditions apply in each case, either
$w_1 =0$ or $w_1 = - \sqrt{3} w_2$ or  $w_1 =  \sqrt{3} w_2$.
\item
Vacuum~R-III requires $\lambda_4=0$ or (in the reducible-triplet framework):
\begin{equation} \label{Eq:lambda4=0,RRF}
4A-2(C+\overline C+D)-E_1+E_2+E_3+E_4=0.
\end{equation}
In this sense, only two of the three minimisation equations are independent.
As a special case of this solution, we can also have $w_S=0$.
This is R-II-3, with $\mu_0^2$ unconstrained.
\end{itemize}

Table~\ref{Table:real} clearly illustrates the point we have made before about the translation
from the reducible-representation framework (RRF) to the irreducible one (IRF). The splitting of the R-I case  
into three cases (a, b, c) would be meaningless due to the $S_3$ symmetry,
if we were only considering $\rho_1$, $\rho_2$
and $\rho_3$. However, in terms of $w_1$, $w_2$ and $w_S$ they appear as different cases.
The consistency of the derivatives with respect to $\hat w_1$ and $\hat w_2$ in this framework is verified since they all have 
$w_S=0$. However now, taking into account the derivative with respect to $w_S$ these solutions must 
obey Eq.~(\ref{4.4}) and each case fulfils this requirement in a different way. We have
$w_2=0$, $w_2 = \sqrt{3} w_1$ and  $w_2 = - \sqrt{3}   w_1$ in the three cases.
Another similar example is case R-II-1. Here the difference is that consistency of the
derivatives with respect to $\hat w_1$ and $\hat w_2$ in the IRF is verified for $w_1=0$, $w_1 = - \sqrt{3} w_2$
and $w_1 = \sqrt{3} w_2$, respectively. Since in this case we do not impose $w_S=0$
the constraint of Eq.~(\ref{4.4})  does not apply. 

The high-scale validity of models based on two of these vacua, namely R-I-1 and R-II-1c, has recently been studied in Ref.~\cite{Chakrabarty:2015kmt}.

\paragraph{Special limits.}
Some of the vacua listed in Table~\ref{Table:real} can be seen as special limits\footnote{The terminology ``special limits'' is not perfect. While a vacuum specification R-X-y is obtained as a special limit of the specification R-X$^\prime$-y$^\prime$, the constraints defining R-X$^\prime$-y$^\prime$ may be a subset of those defining R-X-y. This is analogous to the discussion of real ``origins'' of complex vacua in section~\ref{Sec:transition}.} of another, more general case. These include
\begin{itemize}
\item
R-I-1 is contained in R-II-1a, 1b, 1c for $w_2=0$ (or $x=y$).
\item
R-II-2 is contained in R-II-3 with $w_1=0$ (or $x=y$).
\item
R-II-1a is contained in R-III with $w_1=0$ (or $\rho_1=\rho_2$).
\item
R-II-1a, 1b, 1c, with $\lambda_4=0$, are special cases of R-III.
\item
R-I-2a, 2b, 2c are contained in R-II-3 in the special limits of $w_2=0$, $w_2=\sqrt3 w_1$,
and $w_2=-\sqrt3 w_1$, respectively.
\end{itemize}

\section{Complex vacua}
\label{Sec:ComplexVacua}
\setcounter{equation}{0}

\subsection{The Irreducible-Representation Framework (IRF)}

As a prelude to studying the three complex minimisation equations, we may start with a simpler, linear combination of the last two.
If we in the irreducible-representation framework adopt a convention where $w_S$ is real, and take
\begin{equation} \label{Eq:IRF:notation}
w_1=\hat w_1e^{i\sigma_1}, \quad 
w_2=\hat w_2e^{i\sigma_2},
\end{equation}
with the $\hat {w_i}$ real and non-negative,
then we find the consistency condition (for $w_1\neq0$ and $w_2\neq0$)
\begin{align} \label{Eq:DD-consistence-complex}
&\frac{\partial V}{w_1\partial w_1^\ast}-\frac{\partial V}{w_2\partial w_2^\ast} \nonumber \\
&=2(\lambda_2+\lambda_3)\hat w_1\hat w_2
[\hat w_1^2(e^{i(\sigma_1+\sigma_2)}-e^{i(3\sigma_1-\sigma_2)})
+\hat w_2^2(e^{i(-\sigma_1+3\sigma_2)}-e^{i(\sigma_1+\sigma_2)})] \nonumber \\
&+\lambda_4 w_S\hat w_1[-\hat w_1^2(e^{3i\sigma_1}+2e^{e^{i\sigma_1}})
+\hat w_2^2(3e^{i(\sigma_1+2\sigma_2)}+4e^{i\sigma_1}+2e^{i(-\sigma_1+2\sigma_2)})]
\nonumber \\
&+2\lambda_7w_S^2\hat w_1\hat w_2(e^{i(-\sigma_1+\sigma_2)}-e^{i(\sigma_1-\sigma_2)})=0.
\end{align}
This condition, which is simpler than any of the individual derivatives (\ref{Eq:partial-hS})--(\ref{Eq:partial-h2}), is a {\it necessary}, but not sufficient condition for the vacuum.

If $w_1 = 0$, then equation~(\ref{Eq:partial-h1})  is identically zero, and as a result 
 (\ref{Eq:partial-h1}) and (\ref{Eq:partial-h2}) are automatically consistent. 
The same does not apply to the case $w_2 = 0$ since this case requires  $\lambda_4=0$
for Eq.~(\ref{Eq:partial-h2})  to be satisfied.

The generalisation of equation~(\ref{Eq:lambda4=0,DD}) to the complex case is that the right-hand side of equation~(\ref{Eq:DD-consistence-complex}) be zero. This condition defines a hypersurface in a multidimensional parameter space. Whereas $w_S=0$ and $\lambda_4=0$ are possible solutions in the real case, they are not in the complex case, unless supplemented by additional conditions. For example, if $w_S=0$, we must also have
\begin{equation}
(\lambda_2+\lambda_3)\hat w_1\hat w_2
[\hat w_1^2(1-e^{2i(\sigma_1-\sigma_2)})
+\hat w_2^2(e^{-2i(\sigma_1-\sigma_2)}-1)]=0.
\end{equation}
Splitting this equation into real and imaginary parts, we find
\begin{subequations}
\begin{align}
(\lambda_2+\lambda_3)(\hat w_1-\hat w_2)\{1-\cos[2(\sigma_1-\sigma_2)]\}&=0, \quad \text{and} \\
(\lambda_2+\lambda_3)\sin[2(\sigma_1-\sigma_2)]=0.
\end{align}
\end{subequations}
The solutions are
\begin{subequations}
\begin{align}
\lambda_2+\lambda_3&=0, \quad \text{or} \label{Eq:wS=0,lam_23=0}\\
\sigma_2&=\sigma_1\pm n\pi, \quad n \text{ integer}.
\end{align}
\end{subequations}

\begin{table}[htb]
\caption{Complex vacua. Notation: $\epsilon=1$ and $-1$ for C-III-d and C-III-e, respectively;
$\xi=\sqrt{-3\sin 2\rho_1/\sin2\rho_2}$,
$\psi=\sqrt{[3+3\cos (\rho_2-2 \rho _1)]/(2\cos\rho_2)}$. With the constraints of Table~\ref{Table:Constraints-complex-IV-V} the vacua labelled with an asterisk ($^\ast$) are in fact real.}
\label{Table:complex}
\begin{center}
\begin{tabular}{|c|c|c|}
\hline\hline
& IRF (Irreducible Rep.)& RRF  (Reducible Rep.) \\
\hline
& $w_1,w_2,w_S$ & $\rho_1,\rho_2,\rho_3$  \\
\hline
\hline
C-I-a & $\hat w_1,\pm i\hat w_1,0$ & 
$x, xe^{\pm\frac{2\pi i}{3}}, xe^{\mp\frac{2\pi i}{3}}$ \\
\hline
\hline
C-III-a & $0,\hat w_2e^{i\sigma_2},\hat w_S$ & $y, y, xe^{i\tau}$  \\
\hline
C-III-b & $\pm i\hat w_1,0,\hat w_S$ & $x+iy,x-iy,x$  \\
\hline
C-III-c & $\hat w_1 e^{i\sigma_1},\hat w_2e^{i\sigma_2},0$ 
& $xe^{i\rho}-\frac{y}{2}, -xe^{i\rho}-\frac{y}{2}, y$  \\
\hline
C-III-d,e & $\pm i \hat w_1,\epsilon\hat w_2,\hat{w}_S$ & $xe^{ i\tau},xe^{- i\tau},y$ \\
\hline
C-III-f & $\pm i\hat w_1 ,i\hat w_2,\hat{w}_S$ 
& $re^{i\rho}\pm ix,re^{i\rho}\mp ix,\frac{3}{2}re^{-i\rho}-\frac{1}{2}re^{i\rho}$ \\
\hline
C-III-g & $\pm i\hat w_1,-i\hat w_2,\hat{w}_S$ 
& $re^{-i\rho}\pm ix,re^{-i\rho}\mp ix,\frac{3}{2}re^{i\rho}-\frac{1}{2}re^{-i\rho}$ \\
\hline
C-III-h & $\sqrt{3}\hat w_2 e^{i\sigma_2},\pm\hat w_2 e^{i\sigma_2},\hat{w}_S$ 
& $xe^{i\tau} , y , y$ \\
& & $y, xe^{i\tau},y$ \\
\hline
C-III-i & $\sqrt{\frac{3(1+\tan^2\sigma_1)}{1+9\tan^2\sigma_1}}\hat w_2e^{i\sigma_1},$ 
& $x, ye^{i\tau},ye^{-i\tau}$ \\
& $\pm\hat w_2e^{-i\arctan(3\tan\sigma_1)},\hat w_S$ 
& $ye^{i\tau}, x, ye^{-i\tau}$ \\
\hline
\hline
C-IV-a$^\ast$ & $\hat w_1e^{i\sigma_1},0,\hat w_S$ & $re^{i\rho}+x, -re^{i\rho}+x,x$ \\
\hline
C-IV-b & $\hat w_1,\pm i\hat w_2,\hat w_S$ 
& $re^{i\rho}+x, -re^{-i\rho}+x, -re^{i\rho}+re^{-i\rho}+x$ \\
\hline
C-IV-c & $\sqrt{1+2\cos^2\sigma_2}\hat w_2,$ &  $re^{i\rho}+r\sqrt{3(1+2\cos^2\rho)}+x$, \\
& $\hat w_2e^{i\sigma_2},\hat w_S$ 
& $re^{i\rho}-r\sqrt{3(1+2\cos^2\rho)}+x,-2re^{i\rho}+x$ \\
\hline
C-IV-d$^\ast$ & $\hat w_1e^{i\sigma_1},\pm\hat w_2e^{i\sigma_1},\hat w_S$ & $r_1e^{i\rho}+x, (r_2-r_1)e^{i\rho}+x,-r_2e^{i\rho}+x$ \\
\hline
C-IV-e & $\sqrt{-\frac{\sin 2\sigma_2}{\sin 2\sigma_1}}\hat w_2e^{i\sigma_1},$ & $re^{i\rho_2}+re^{i\rho_1}\xi+x, re^{i\rho_2}-re^{i\rho_1}\xi+x,$  \\
& $\hat w_2e^{i\sigma_2},\hat w_S$ & $-2re^{i\rho_2}+x$ \\
\hline
C-IV-f & $\sqrt{2+\frac{\cos \left(\sigma _1-2 \sigma _2\right)}{\cos\sigma_1}}\hat w_2e^{i\sigma_1},$ & $re^{i\rho_1}+re^{i\rho_2}\psi+x$, \\
& $\hat w_2e^{i\sigma_2},\hat w_S$ &$re^{i\rho_1}-re^{i\rho_2}\psi+x, -2re^{i\rho_1}+x$ \\
\hline
\hline
C-V$^\ast$ & $\hat w_1e^{i\sigma_1},\hat w_2e^{i\sigma_2},\hat w_S$ & $xe^{i\tau_1},ye^{i\tau_2},z$ \\
\hline
\end{tabular}
\end{center}
\end{table}

Likewise, the condition $\lambda_4=0$ must be supplemented by 
\begin{align} \label{Eq:constraint-for-lambda4=0}
&(\lambda_2+\lambda_3)e^{i(\sigma_1+\sigma_2)}[\hat w_1^2(1-e^{2i(\sigma_1-\sigma_2)})
+\hat w_2^2(e^{-2i(\sigma_1-\sigma_2)}-1)] \nonumber \\
&+\lambda_7 w_S^2(e^{-i(\sigma_1-\sigma_2)}-e^{i(\sigma_1-\sigma_2)})=0.
\end{align}
Splitting this equation into real and imaginary parts, we find:
\begin{subequations} \label{Eq:lam4=0}
\begin{gather}
(\lambda_2+\lambda_3)\sin(\sigma_1-\sigma_2)(\sin2\sigma_1\hat w_1^2
+\sin2\sigma_2\hat w_2^2)=0, 
\quad \text{and} \\
[(\lambda_2+\lambda_3)(\cos2\sigma_1\hat w_1^2+\cos2\sigma_2\hat w_2^2)+\lambda_7\hat w_S^2]\sin(\sigma_1-\sigma_2)=0.
\end{gather}
\end{subequations}
Furthermore, Eq.~(\ref{Eq:constraint-for-lambda4=0}) is obviously satisfied for
\begin{equation} \label{Eq:lam7=lam23=0}
\lambda_2+\lambda_3=0, \quad \text{and } \lambda_7=0.
\end{equation}
On the other hand, for
\begin{equation} \label{Eq:tau-relation}
\sigma_2=\sigma_1\pm n\pi, \quad n \text{ integer},
\end{equation}
$\lambda_2$, $\lambda_3$ and $\lambda_7$ are not constrained by equation~(\ref{Eq:constraint-for-lambda4=0}).

Finally, the real-case consistency condition $\hat w_1^2=3\hat w_2^2$ would in the complex case have to be supplemented with the above phase constraint (\ref{Eq:tau-relation}).

We find the solutions given in Table~\ref{Table:complex}. 
The table is organised as follows. From left to right, the first column gives the name of the vacuum, the second gives the specification in the irreducible-representation
framework (IRF), and the third gives its translation to the reducible-representation framework (RRF). 

The need to introduce the parameter $\epsilon$ in Table~\ref{Table:complex}
results from the definitions given by Eq.~(\ref{Eq:IRF:notation}), where
the $\hat {w_i}$ , $i= 1,2$ are chosen to be non-negative.
Naively one might expect the number of constraints to be equal
to the number of free parameters of the solution. This is not the case,
as can be illustrated by considering  Eq.~(\ref{Eq:IRF-w_S}) in the limit $w_S = 0$.

In the complex case there is a richer structure 
of possible vacua and  again we have solutions that are similar when specified in terms
of $\rho$'s but not in terms of $w$'s.  Furthermore, in some cases solutions which 
can be described in an elegant  way in one of the frameworks do not look so elegant
in the other or may fall  into a particular case of a more general one already given in that framework. 
As an illustration of the first remark let us consider the 
solution $(\rho_1, \rho_2, \rho_3) = x(1,1,e^{i\tau})$, this is a special case of C-III-a, with $y=x$,
however the same solution after a permutation becomes a special case of C-III-h. On the other 
hand, the solution $(\rho_1, \rho_2, \rho_3) = x(1,e^{i\tau},e^{-i\tau})$ is a special case of C-III-i,
with solution C-I-a a special case of this one again.

\begin{table}[htb]
\caption{Constraints on complex vacua. Notation: $\epsilon=1$ and $-1$ for C-III-d and C-III-e, respectively.
Where two possible signs ($\pm$ or $\mp$) are given, they correspond to those of Table~\ref{Table:complex}. Here, $\lambda_b$ is defined in Eq.~(\ref{Eq:lambda_ab}).}
\label{Table:Constraints-complex-I-III}
\begin{center}
\begin{tabular}{|c|c|c|}
\hline\hline
Vacuum &  Constraints \\
\hline
\hline
C-I-a & $\mu _1^2=-2\left(\lambda _1-\lambda _2\right) \hat{w}_1^2$ \\
\hline
\hline
C-III-a & $\mu _0^2=- \frac{1}{2} \lambda_b \hat{w}_2^2-\lambda _8 \hat{w}_S^2$,\\
& $\mu _1^2=-\left(\lambda _1+\lambda _3\right) \hat{w}_2^2- \frac{1}{2}  \left( \lambda _b-8 \cos^2 \sigma _2 \lambda _7\right)\hat{w}_S^2$, \\
& $\lambda _4= \frac{4 \cos \sigma _2 \hat{w}_S }{\hat{w}_2}\lambda _7$ \\
\hline
C-III-b & $\mu _0^2=-\frac{1}{2} \lambda_b \hat{w}_1^2-\lambda _8 \hat{w}_S^2$,\\
& $\mu _1^2=- \left( \lambda _1+ \lambda _3\right) \hat{w}_1^2-\frac{1}{2} \lambda_b \hat{w}_S^2$, \\
& $\lambda _4=0$ \\
\hline
C-III-c & $\mu_1^2=-(\lambda_1+\lambda_3)(\hat{w}_1^2+\hat{w}_2^2)$,\\
& $\lambda_2+\lambda_3=0, \lambda_4=0$ \\
\hline
C-III-d,e & $\mu _0^2=\left(\lambda _2+\lambda _3\right)\frac{ ( \hat{w}_1^2-\hat{w}_2^2)^2}{\hat{w}_S^2}
-\epsilon\lambda _4\frac{( \hat{w}_1^2-\hat{w}_2^2)( \hat{w}_1^2-3\hat{w}_2^2)}{4 \hat{w}_2 \hat{w}_S}$\\
& $-\frac{1}{2}\left(\lambda _5+\lambda _6\right) ( \hat{w}_1^2+\hat{w}_2^2) -\lambda _8 \hat{w}_S^2$, \\
& $\mu _1^2=- \left( \lambda _1- \lambda _2\right) (\hat{w}_1^2+\hat{w}_2^2)
-\epsilon\lambda _4\frac{ \hat{w}_S( \hat{w}_1^2- \hat{w}_2^2)}{4 \hat{w}_2}
-\frac{1}{2} \left( \lambda _5+ \lambda _6\right) \hat{w}_S^2 $,\\
 & $\lambda _7=  \frac{\hat{w}_1^2-\hat{w}_2^2}{\hat{w}_S^2}(\lambda _2+\lambda _3)
 - \epsilon\frac{ (\hat{w}_1^2-5\hat{w}_2^2)}{4 \hat{w}_2\hat{w}_S}\lambda _4$\\
\hline
C-III-f,g & $\mu _0^2= -\frac{1}{2} \lambda_b \left(\hat{w}_1^2+\hat{w}_2^2\right)-\lambda _8 \hat{w}_S^2,$\\
& $\mu _1^2= -\left(\lambda _1+\lambda _3\right) \left(\hat{w}_1^2+\hat{w}_2^2\right)-\frac{1}{2} \lambda_b \hat{w}_S^2, \lambda_4=0$ \\
\hline
C-III-h & $\mu _0^2= -2 \lambda_b \hat{w}_2^2-\lambda _8 \hat{w}_S^2,$\\
& $\mu _1^2=-4 \left(\lambda _1+\lambda _3\right) \hat{w}_2^2 -\frac{1}{2}  \left(\lambda _b-8 \cos^2 \sigma _2 \lambda _7\right)\hat{w}_S^2,$\\
& $\lambda _4= \mp\frac{2  \cos \sigma _2 \hat{w}_S}{\hat{w}_2}\lambda _7$ \\
\hline
C-III-i & $\mu _0^2=\frac{16 \left(1-3\tan^2\sigma_1\right){}^2}{ \left(1+9\tan^2\sigma_1\right){}^2}(\lambda_2+\lambda_3)\frac{\hat{w}_2^4}{\hat{w}_S^2}
\pm\frac{6  \left(1-\tan^2\sigma_1\right) (1-3\tan^2\sigma_1)}{(1+9 \tan ^2\sigma _1)^\frac{3}{2} }\lambda_4\frac{\hat{w}_2^3}{\hat{w}_S}$\\
& $-\frac{2(1+3  \tan^2\sigma_1)}{1+9 \tan ^2\sigma _1}(\lambda_5+\lambda_6)\hat{w}_2^2-\lambda_8\hat{w}_S^2$, \\
&$\mu _1^2=-\frac{4(1+3  \tan^2\sigma_1)}{1+9 \tan ^2\sigma _1}(\lambda_1-\lambda_2)\hat{w}_2^2\mp\frac{ \left(1-3 \tan ^2\sigma _1\right)}{2 \sqrt{1+9 \tan ^2\sigma _1}}\lambda_4 \hat{w}_2 \hat{w}_S$\\
& $-\frac{1}{2}(\lambda_5+\lambda_6)\hat{w}_S^2$, \\
& $\lambda_7=-\frac{4  \left(1-3 \tan ^2\sigma _1\right)\hat{w}_2^2}{ \left(1+9 \tan ^2\sigma _1\right)\hat{w}_S^2}(\lambda_2+\lambda_3)\mp\frac{ \left(5-3 \tan ^2\sigma _1\right)\hat{w}_2}{2  \sqrt{1+9 \tan ^2\sigma _1}\hat{w}_S}\lambda_4$ \\
\hline
\end{tabular}
\end{center}
\end{table}

In Tables~\ref{Table:Constraints-complex-I-III} and \ref{Table:Constraints-complex-IV-V} we list the conditions on the potential parameters, in the irreducible-representation framework.

\begin{table}[htb]
\caption{Constraints on complex vacua, continued. The vacua labelled with an asterisk ($^\ast$) are in fact real.}
\label{Table:Constraints-complex-IV-V}
\begin{center}
\begin{tabular}{|c|c|c|}
\hline\hline
Vacuum &  Constraints \\
\hline
\hline
C-IV-a$^\ast$ & $\mu _0^2=-\frac{1}{2} \left(\lambda _5+\lambda _6\right) \hat{w}_1^2-\lambda _8 \hat{w}_S^2$,\\
& $\mu _1^2=-\left( \lambda _1+ \lambda _3\right) \hat{w}_1^2-\frac{1}{2} \left(\lambda _5+\lambda _6\right) \hat{w}_S^2$, \\
&$\lambda _4=0, \lambda _7=0$ \\
\hline
C-IV-b & $\mu _0^2=\left(\lambda _2+\lambda _3\right)\frac{ \left(\hat{w}_1^2-\hat{w}_2^2\right){}^2}{\hat{w}_S^2}
-\frac{1}{2} \left(\lambda _5+\lambda _6\right) \left(\hat{w}_1^2+\hat{w}_2^2\right) -\lambda _8 \hat{w}_S^2$,\\
& $\mu _1^2=-\left(\lambda _1-\lambda _2\right) \left(\hat{w}_1^2+\hat{w}_2^2\right) -\frac{1}{2} \left(\lambda _5+\lambda _6\right) \hat{w}_S^2$, \\
&$\lambda _4=0, \lambda _7= -\frac{ \left(\hat{w}_1^2-\hat{w}_2^2\right)}{\hat{w}_S^2}\left(\lambda _2+\lambda _3\right)$ \\
\hline
C-IV-c 
& $\mu _0^2=2 \cos ^2\sigma _2 \left(1+\cos^2 \sigma _2\right)\left(\lambda _2+\lambda _3\right)\frac{\hat{w}_2^4}{\hat{w}_S^2}$\\
& 
$-\left(1+\cos^2 \sigma _2\right)\left(\lambda _5+\lambda _6\right) \hat{w}_2^2 -\lambda _8 \hat{w}_S^2 $,\\
& 
$\mu _1^2=-\left[ 2 \left(1+\cos^2 \sigma _2\right)\lambda _1- \left(2+3 \cos^2 \sigma _2\right)\lambda _2-  \cos ^2\sigma _2\lambda _3\right] \hat{w}_2^2 $\nonumber\\
& $ -\frac{1}{2} \left(\lambda _5+\lambda _6\right) \hat{w}_S^2$, \\
&$\lambda _4= -\frac{2   \cos \sigma _2 \hat{w}_2}{\hat{w}_S}\left(\lambda _2+\lambda _3\right),\lambda _7= \frac{  \cos ^2\sigma _2 \hat{w}_2^2}{\hat{w}_S^2}\left(\lambda _2+\lambda _3\right)$ \\
\hline
C-IV-d$^\ast$ & $\mu _0^2=-\frac{1}{2} \left(\lambda _5+\lambda _6\right) (\hat{w}_1^2+\hat{w}_2^2)-\lambda _8 \hat{w}_S^2$,\\
& $\mu _1^2=-\left( \lambda _1+ \lambda _3\right)  (\hat{w}_1^2+\hat{w}_2^2)-\frac{1}{2} \left(\lambda _5+\lambda _6\right) \hat{w}_S^2$, \\
&$\lambda _4=0, \lambda _7=0$ \\
\hline
C-IV-e & $\mu _0^2=\frac{ \sin ^2\left(2 \left(\sigma _1-\sigma _2\right)\right)}{ \sin ^2\left(2 \sigma _1\right)}\left(\lambda _2+\lambda _3\right) \frac{\hat{w}_2^4}{\hat{w}_S^2}$ \\
& $ -\frac{1}{2}\left(1-\frac{\sin 2 \sigma _2}{ \sin 2 \sigma _1}\right) \left(\lambda _5+\lambda _6\right) \hat{w}_2^2-\lambda _8 \hat{w}_S^2$, \\
&$\mu _1^2= - \left(1-\frac{\sin 2 \sigma _2}{ \sin 2 \sigma _1}\right) \left(\lambda _1-\lambda _2\right) \hat{w}_2^2-\frac{1}{2}\left(\lambda _5+\lambda _6\right) \hat{w}_S^2$,\\
& $\lambda_4=0, \lambda _7= -\frac{  \sin \left(2 \left(\sigma _1-\sigma _2\right)\right) \hat{w}_2^2}{\sin 2 \sigma _1\hat{w}_S^2}\left(\lambda _2+\lambda _3\right)$ \\
\hline
C-IV-f & $\mu _0^2=-\frac{ \left(\cos \left(\sigma _1-2 \sigma _2\right) +3\cos\sigma_1\right)\cos(\sigma_2-\sigma_1)}{2\cos^2\sigma_1}\lambda _4\frac{\hat{w}_2^3}{ \hat{w}_S}$\\
& $ - \frac{ \cos \left(\sigma _1-2 \sigma _2\right) +3\cos\sigma_1}{2\cos\sigma_1}\left(\lambda _5+\lambda _6\right) \hat{w}_2^2-\lambda _8 \hat{w}_S^2$, \\
&$\mu _1^2=- \frac{ \cos \left(\sigma _1-2 \sigma _2\right) +3\cos\sigma_1}{\cos\sigma_1}\left(\lambda _1+\lambda _3\right) \hat{w}_2^2  $ \\
& $-  \frac{3 \cos 2 \sigma _1+2 \cos \left(2 \left(\sigma _1-\sigma _2\right)\right)+\cos 2 \sigma _2+4}
{4\cos \left(\sigma _1-\sigma _2\right) \cos \sigma _1} \lambda _4 \hat{w}_2 \hat{w}_S 
- \frac{1}{2}\left(\lambda _5+\lambda _6\right) \hat{w}_S^2$,\\
& $\lambda_2+\lambda _3=-\frac{ \cos \sigma _1 \hat{w}_S}{2 \cos (\sigma_2-\sigma_1)\hat{w}_2}\lambda _4 , \lambda _7=-\frac{  \cos (\sigma_2-\sigma_1)\hat{w}_2}{2 \cos\sigma_1\hat{w}_S}\lambda _4$ \\
\hline
C-V$^\ast$ & 
$\mu _0^2=-\frac{1}{2} \left(\lambda _5+\lambda _6\right) (\hat{w}_1^2+\hat{w}_2^2) -\lambda _8 \hat{w}_S^2$,\\
& $\mu _1^2= -\left(\lambda _1+\lambda _3\right) (\hat{w}_1^2+\hat{w}_2^2)-\frac{1}{2} \left(\lambda _5+\lambda _6\right) \hat{w}_S^2$, \\
&$\lambda _2+\lambda _3=0,\lambda _4=0,\lambda _7= 0$ \\
\hline
\end{tabular}
\end{center}
\end{table}

The vacuum C-III-c (see Tables~\ref{Table:complex} and \ref{Table:Constraints-complex-I-III}) falls in the category satisfying Eq.~(\ref{Eq:wS=0,lam_23=0}).
Examples of the constraints (\ref{Eq:lam4=0}) are to be found in the vacua C-IV-b and C-IV-e (see Tables~\ref{Table:complex} and \ref{Table:Constraints-complex-IV-V}). 
The constraints  (\ref{Eq:lam7=lam23=0}) apply to the vacuum C-V (see Tables~\ref{Table:complex} and \ref{Table:Constraints-complex-IV-V}).

The case C-III-c of Table~\ref{Table:complex} is very interesting.
It is a solution with $\hat w_S=0$ and $\lambda_4=0$ with the
additional constraint $\lambda_2+\lambda_3=0$, allowing for a 
relative phase between the vevs of $h_1$ and $h_2$. The fact that 
$\hat w_S$ is equal to zero suggests that this vacuum may provide a
viable dark matter candidate.
This vacuum can be specified in terms of two non-zero moduli and one single
phase. Once we replace $\lambda_2$ in terms of $\lambda_3$, the
Lagrangian is left with only one term that is sensitive to
the relative phase between $h_1$ and $h_2$, to wit the term in $\lambda_7$.
The fact that the moduli of the two vevs are different might lead one to think 
that this vacuum violates CP spontaneously. However, in section 8 we show
that this is not the case.

Cases C-IV-a, C-IV-d  and C-V are listed in Table 2 for
completeness and to allow for an enlightening discussion.         
Once one takes into consideration the constraints given in
Table 4 they become real. 

Solution C-IV-d is more general than solution C-IV-a
and reduces to  C-IV-a once we fix $w_2 = 0$, so it suffices to discuss C-IV-d.
Both of these require $\lambda_4 = 0$ and $\lambda_7 =0$, and as a result
the potential acquires  symmetry for the transformation of
$h_1$, $h_2$ and $h_S$ under a unitary transformation of the form 
$U = \mbox{diag}(e^{i \tau}, e^{i \tau}, 1)$
which allows to remove the phase $\sigma_1$ from the vacuum, making it real.

At first glance case C-V looks like the most general case,
however we are assuming that it does not fall into any of the previous cases,
so, as a result, full generality  requires $\lambda_2 + \lambda_3 =0$,
$\lambda_4 =0$ and $\lambda_7 =0$ and there is no term in the potential
sensitive to independent rephasing of each of the $h$ fields. As a
result any phase in the vevs can be rotated  away.
Under these circumstances, it is equivalent to a real set of vacua.
 
There are, in particular, two possible complex vacua that have been
discussed previously in the literature. One of them is:
\begin{equation}
\hat{w} e^{i \sigma}, \quad \hat{w} e^{- i \sigma}, \quad \hat{w_S},
\label{PS}
\end{equation}
by Pakvasa and Sugawara \cite{Pakvasa:1977in}. We shall refer to this as the PS vacuum,
assuming $\hat w\neq 0$ and $\hat w_S\neq 0$.
There is also a solution given by Ivanov and Nishi \cite{Ivanov:2014doa}
\begin{equation}
\hat{w} e^{i \sigma}, \quad \hat{w} e^{ i \sigma}, \quad \hat{w_S},
\label{IN}
\end{equation}
which we shall refer to as the IN vacuum, assuming again $\hat w\neq0$ and $\hat w_S\neq0$.
By imposing the minimisation conditions it can be checked that both of  these solutions
require  $\lambda_4$  equal to zero, corresponding to the $SO(2)$ symmetry of the
potential given by Eq.~(\ref{so2}). It is clear that Eq.~(\ref{IN}) does not break this
symmetry spontaneously.

The PS vacuum specified by Eq.~(\ref{PS}) is only consistent for the following choices: 
\begin{alignat}{3}
&\text{PS-a}: &\quad &\lambda_4 =0 , \quad \text{and }
\sigma=\pm \pi/2,
&\quad &\text{included in case C-III-f,g},\\
&\text{PS-b}: &\quad &\lambda_4 =0 , \quad \text{and }\lambda_7 = -2\cos2\sigma\frac{\hat w^2}{\hat w_S^2}(\lambda_2+\lambda_3),
&\quad &
\begin{minipage}{5cm}
included in case C-IV-e \\
with $\sigma_2=-\sigma_1.$
\end{minipage}
\end{alignat}
As a special case of PS-b, we can have $\lambda_7=0$ and $\sigma=\pi/4$ or $3\pi/4$, with $(\lambda_2+\lambda_3)$ unconstrained.

The IN vacuum specified by  Eq.~(\ref{IN}) is only consistent if either of the following
two sets of conditions is verified:
\begin{alignat}{3}
&\text{IN-a}: &\quad &\lambda_4 =0 , \quad \text{and }\sigma = \pm\pi/2,
&\quad &\text{included in C-III-f,g}, \label{INa} \\
&\text{IN-b}: &\quad &\lambda_4 =0 , \quad \text{and }\lambda_7=0,
&\quad &\text{included in case C-IV-d}. \label{INb}
\end{alignat} 
As discussed above, the conditions listed under  IN-b lead to a real vacuum since 
they allow for the common phase of  $w_1$ and $w_2$ to be rotated away.  

\paragraph{Special limits.}
Some of the vacua listed in Table~\ref{Table:complex} can be seen as special limits of another, more general case. These include
\begin{itemize}
\item
C-III-a and C-III-h are equivalent in terms of the RRF.
\item
C-III-b, for $\lambda_7=0$ becomes real and falls into C-IV-a.
\item
C-IV-a is contained in C-IV-d with $\hat w_2=0$.
\item
Solution C-IV-e reduces to C-IV-b for $\sigma_2=\sigma_1\pm\pi/2$, in the limit $\sigma_1\to0$.
\item
C-IV-c is contained in C-IV-f for $\sigma_1=0$.
\end{itemize}

\subsection{The Reducible-Representation Framework (RRF)}

Below follow further general comments on the different vacua, as well as some constraints on the parameters in the reducible-representation framework:
\begin{itemize}
\item
Vacuum C-I-a requires
\begin{equation} \label{Eq:C-I-a-RRF}
2\lambda+\gamma=x^2[2A+2C+2\overline C-D-2E_1-E_2-E_3+2E_4].
\end{equation}
\item
Vacuum C-III-h allows for a particular realisation in terms of the RRF with $y=x$, of the form 
\begin{equation}
\text{C-RRF-a:}\qquad (\rho_1,\rho_2,\rho_3)=x(1,e^{i\tau},e^{i\tau}). 
\end{equation}
Here, we have applied an overall phase rotation to the first solution presented in Table~\ref{Table:complex} for this vacuum, thus complying with the notation defined by Eq.~(\ref{Eq:RRF-vacuum}).
Depending on where we put the phase, we may have two different translations in terms of the IRF: one of them in C-III-h and the other in C-III-a. This leads to three independent minimisation conditions (as specified by the label III), two ``radial'' ones from $\partial V/\partial v_1=0$ and $\partial V/\partial v_2=0$ and one ``angular'' one from $\partial V/\partial\tau=0$. The two radial equations are quadratic in $\cos\tau$, whereas the third is linear. With $x$ and $\tau$ as input, these equations allow to constrain three parameters of the potential. Alternatively, one may remove terms quadratic in $\cos\tau$ by forming linear combinations of the two radial equations.
With the rescaling
\begin{equation} \label{Eq:rescale-bilinears}
\lambda^\prime=\lambda/x^2, \quad
\gamma^\prime=\gamma/x^2,
\end{equation}
we find two radial equations
\begin{subequations}
\begin{align}
\cos\tau
&=\frac{-4D-4E_1+E_2+E_3-E_4-2\gamma^\prime}
{2(2D+E_4)}, \\
&=\frac{-A-C-\overline C-D-2E_1-\gamma^\prime+\lambda^\prime}
{E_2+E_3+E_4},
\end{align}
\end{subequations}
and the angular equation
\begin{equation}
\cos\tau=\frac{-2E_1-E_2-E_3-E_4-2\gamma^\prime}
{2(2D+E_4)}.
\end{equation}
Consistency of the first and the third expressions leads to
\begin{equation} \label{Eq:quartic-cond}
2D+E_1-E_2-E_3=0,
\end{equation}
which corresponds to
\begin{equation} \label{Eq:lambda_4_vs_lambda_7}
\lambda_4=\sqrt{2}\lambda_7.
\end{equation}
Consistency of the second and third equation leads to:
\begin{equation}
(2A+2C+2\overline C-E_4)(2D+E_4)-3E_1^2-2(2D+E_4)\lambda^\prime-2E_1\gamma^\prime=0.
\end{equation}
Invoking Eq.~(\ref{Eq:rescale-bilinears}), we may solve for $x^2$:
\begin{equation}
x^2=\frac{2(2D+E_4)\lambda+2E_1\gamma}{(2A+2C+2\overline C-E_4)(2D+E_4)-3E_1^2}.
\end{equation}
The conditions
\begin{equation}
x^2>0 \quad \text{and} \quad
|\cos\tau|\leq1
\end{equation}
will further constrain the potential parameters for this particular vacuum.

As presented in Table~\ref{Table:complex}, Case C-III-h illustrates, once again, the fact
that trivial permutations of the reducible triplet lead to different constraints for the IRF.

\item
The cases C-III-d, C-III-e and C-III-i, when presented in the RRF, in the limit $y=x$, become
\begin{equation}
\text{C-RRF-b:}\qquad (\rho_1,\rho_2,\rho_3)=x(1,e^{i\tau},e^{-i\tau}),
\end{equation}
and permutations\footnote{Vacuum C-I-a is of course a special case of this one, with $\tau=2\pi/3$.}. These three IRF cases merge into one RRF case.

There are three minimisation conditions, involving $x^2$ and $\cos\tau$. The constraint from the minimisation with respect to $\tau$ can be expressed as a $\cos\tau$-dependent relation among the quartic terms:
\begin{equation} \label{Eq:E_23_vs_others}
E_2+E_3=(4\cos\tau-2)D+E_1+2(1-\cos\tau)E_4,
\end{equation}
whereas the others can be solved for $\gamma$ and $\lambda$. Making use of Eq.~(\ref{Eq:E_23_vs_others}), these take the form
\begin{align}
\gamma&=\frac{x^2}{2}[(2-8\cos^2\tau)D-3E_1+(1-4\cos\tau)E_4], \label{Eq:C-RRF-b-gamma} \\
\lambda&=\frac{x^2}{2}[2(A+C+\overline C)-4\cos\tau(1-\cos\tau)D
+(2\cos^2\tau+2\cos\tau-1)E_1 \nonumber \\
&\quad -(2\cos^2\tau-2\cos\tau+1)E_4].
\label{Eq:C-RRF-b-lambda}
\end{align}
The three constraints (minimisation conditions) of Table~\ref{Table:Constraints-complex-I-III}
will for C-III-d, C-III-e, and C-III-i take forms
equivalent to these Eqs.~(\ref{Eq:E_23_vs_others})--(\ref{Eq:C-RRF-b-lambda}). Equations (\ref{Eq:C-RRF-b-gamma}) and (\ref{Eq:C-RRF-b-lambda}) can be solved for $x^2$, but the two solutions impose a $\cos\tau$-dependent consistency condition on the coefficients of the potential, given by Eq.~(\ref{Eq:E_23_vs_others}).

\item
The vacua C-RRF-a and C-RRF-b have the same form as the PS and IN vacua of Eqs.~(\ref{PS})--(\ref{INb}). However, it must be stressed that they refer to the fields of the reducible-representation framework.
\end{itemize}

\section{Complex vacua vs real vacua}
\label{Sec:transition}
\setcounter{equation}{0}

The complex vacua, which are specified by three moduli and two (relative) phases are found as solutions of five conditions, whereas the real vacua are found as solutions of three. The following questions then arise: Can the complex vacua be seen as generalisations of the real ones? Are the conditions on the moduli compatible with those for one or more of the real vacua? Are these more restrictive, or less restrictive?

In order to discuss how a complex vacuum may be related to a real one, let us introduce some notations. Let us denote by ${\cal C}$(C-X-y) the set of constraints (such as given in Tables~\ref{Table:Constraints-complex-I-III} and \ref{Table:Constraints-complex-IV-V}) satisfied by a particular complex vacuum. Likewise, we let the real vacuum R-X$^\prime$-y$^\prime$ satisfy the constraints ${\cal C}$(R-X$^\prime$-y$^\prime$) (see Table~\ref{Table:real}).
Then, we may consider a real vacuum R-X$^\prime$-y$^\prime$ the ``origin'' of a particular complex vacuum C-X-y if the following two conditions are satisfied:
\begin{itemize}
\item
the C-X-y specification, in an existing real limit for the vacuum (there may be an ambiguity of sign) coincides with that of R-X$^\prime$-y$^\prime$, and
\item
the constraints are compatible,
\begin{equation}
{\cal C}(\text{C-X-y}) \subset {\cal C}(\text{R-X$^\prime$-y$^\prime$}) .
\end{equation}
\end{itemize}
The latter condition is important due to the fact that the transition to a real vacuum is not always possible.

We list in Table~\ref{Table:transitions} the real vacua satisfying these two requirements. 

\begin{table}[htb]
\caption{Transitions from complex to real vacua in the IRF. The vacua labelled with an asterisk ($^\ast$) were shown to be real.}
\label{Table:transitions}
\begin{center}
\begin{tabular}{|c|c|c|c|}
\hline\hline
Complex & Real ``origin'' \\
\hline
\hline
C-I-a & none \\
\hline
\hline
C-III-a & R-II-1a \\
\hline
C-III-b & none \\
\hline
C-III-c & R-I-2a,2b,2c, R-II-3 \\
\hline
C-III-d,e & none \\
\hline
C-III-f & none \\
\hline
C-III-g & none \\
\hline
C-III-h & R-II-1b,1c \\
\hline
C-III-i & R-II-1b,1c \\
\hline
\hline
C-IV-a$^\ast$ & R-III \\
\hline
C-IV-b & none  \\
\hline
C-IV-c & R-II-1b,1c \\
\hline
C-IV-d$^\ast$ & R-III \\
\hline
C-IV-e & none  \\
\hline
C-IV-f & R-II-1b,1c \\
\hline
\hline
C-V$^\ast$ & R-III \\
\hline
\hline
\end{tabular}
\end{center}
\end{table}

In the following subsection, we study a particular example, how a complex vacuum is related to a real one.

\subsection{Transition from Vacua~R-II-1b,1c to Vacuum~C-III-h}

An important difference between the constraints of the vacua R-II-1b,1c and C-III-h is that in the former case, the potential parameters $\lambda_4$ and $\lambda_7$ are free, i.e., they are uncorrelated. For the vacuum C-III-h, on the other hand, they are correlated as (see Table~\ref{Table:Constraints-complex-I-III})
\begin{equation} \label{Eq:lines-lambda4_lambda7}
\lambda _4= \mp\frac{2  \cos \sigma _2 \hat{w}_S}{\hat{w}_2}\lambda _7.
\end{equation}
Modulo positivity and other physical constraints, the whole $\lambda_4$--$\lambda_7$-plane is available for the real vacuua R-II-1b,1c, whereas only the lines defined by Eq.~(\ref{Eq:lines-lambda4_lambda7}) are available for an ``extension'' to a complex vacuum C-III-h. This holds even for infinitesimal phases, i.e., $\cos\tau_2\to\pm1$.

The following question arises: Under what conditions is the complex vacuum deeper? It turns out that the difference can be expressed as being proportional to $\lambda_4$ or $\lambda_7$:
\begin{equation}
\Delta V\equiv V(\text{R-II-1b,1c})-V(\text{C-III-h})
=-4\lambda_7\hat w_2^2\hat w_S^2(1\mp\cos\sigma_2)^2.
\end{equation}
The question of relating complex vacua to real ones is relevant for the discussion of global minima 
\cite{Barroso:2006pa,EmmanuelCosta:2007zz}
as well as to understand the possible correlations of different parameters of the potential.

\section{The case of $\lambda_4=0$}
\label{Sec:lambda4}
\setcounter{equation}{0}
As mentioned in section~\ref{Sec:DasDey}, in the case of $\lambda_4=0$
the potential has an additional, continuous $SO(2)$ symmetry.
This case was dismissed by Derman \cite{Derman:1978rx}, as being ``un-natural''.
This was due to the fact that this condition, when expressed in terms of the
parameters of the potential written by Derman, given by Eqs.~(\ref{29ab}),
acquires the form given by Eq.~(\ref{Eq:lambda4=0,RRF}),
which is not instructive and the resulting symmetry is not apparent.
Spontaneous breaking of this $SO(2)$ symmetry leads to
massless particles. In this case, one way to promote this to a viable model
is to break  this symmetry softly, by adding a term to the bilinear part of the potential:
\begin{equation} 
V=V_2 + V_2^\prime + V_4,
\end{equation}
with $V_2$ and $V_4$ as defined by equations~(\ref{Eq:V-DasDey}). Choosing
\begin{equation} \label{Eq:V_soft}
V_2^\prime=\frac{1}{2}\nu^2(h_2^\dagger h_1 + h_1^\dagger h_2),
\end{equation}
the minimisation conditions (\ref{Eq:partial-hS})--(\ref{Eq:partial-h2}) will now become
\begin{align} \label{Eq:partial-hS-soft}
\frac{\partial V}{\partial w_S^\ast}&=
\frac{1}{2}w_S\mu_0^2
+\frac{1}{4}w_S(|w_1|^2+|w_2|^2)(\lambda_5+\lambda_6)
 \nonumber \\
&+\frac{1}{4}w_S^\ast(w_1^2+w_2^2)\lambda_7
+\frac{1}{2}w_S^\ast w_S^2\lambda_8=0, \\
\label{Eq:partial-h1-soft}
\frac{\partial V}{\partial w_1^\ast}&=\frac{1}{2}w_1\mu_1^2
+\frac{1}{2}w_2\nu^2
+\frac{1}{2}w_1(|w_1|^2+|w_2|^2)\lambda_1
+\frac{1}{2}w_2(w_1^\ast w_2-w_1 w_2^\ast)\lambda_2 \nonumber \\
&+\frac{1}{2}w_1^\ast(w_1^2+w_2^2)\lambda_3
+\frac{1}{4}w_1|w_S|^2(\lambda_5+\lambda_6)
+\frac{1}{2}w_1^\ast w_S^2\lambda_7=0, \\
\label{Eq:partial-h2-soft}
\frac{\partial V}{\partial w_2^\ast}&=\frac{1}{2}w_2\mu_1^2
+\frac{1}{2}w_1\nu^2
+\frac{1}{2}w_2(|w_1|^2+|w_2|^2)\lambda_1
-\frac{1}{2}w_1(w_1^\ast w_2-w_1 w_2^\ast)\lambda_2 \nonumber \\
&+\frac{1}{2}w_2^\ast(w_1^2+w_2^2)\lambda_3
+\frac{1}{4}w_2|w_S|^2(\lambda_5+\lambda_6)
+\frac{1}{2}w_2^\ast w_S^2\lambda_7=0.
\end{align}

With these new conditions there will be some changes in the vacuum solutions.
Notice that such a term also softly breaks the discrete symmetries $h_1$ 
going into $- h_1$ and $h_2$ going into $- h_2$. 
Another possible choice for a term breaking softly
the $SO(2)$ symmetry is:
\begin{equation}
V_2^\prime=\frac{1}{2}\mu_2^2(h_1^\dagger h_1 - h_2^\dagger h_2).
\end{equation}
Soft breaking terms involving $h_S$ and one $h_i$ are not consistent with
$\lambda_4=0$.
It was shown, in the context of two Higgs doublet models with a discrete
symmetry, that CP can only be violated spontaneously once a soft symmetry breaking term is 
added to the potential \cite{Branco:1985aq}.
Soft  breaking of the $S_3$ symmetry of the scalar potential has been applied
in \cite{Koide:2005ep} in order to obtain a  special relation among the vevs of the 
three doublets that would allow to account  for the observed charged lepton masses.

An important implication of the type of vacuum solution
and of the corresponding allowed region of parameter space is the
resulting different possible spectra for the physical scalars. 

\section{Spontaneous CP violation}
\label{Sec:CPV}
\setcounter{equation}{0}
The $S_3$-symmetric potential offers a very rich phenomenology, and can accommodate a
variety of physical situations, as outlined in
sections~\ref{Sec:RealVacua} and \ref{Sec:ComplexVacua}, where we classified the different vacua. 

We assumed, for simplicity,  that all parameters of the potential are real.
Therefore our discussion is done in the framework of explicit CP
conservation. This raises the question of whether or not CP can be violated
spontaneously. For that purpose we can inspect the list of complex solutions
presented in Table~\ref{Table:complex}. CP can only be spontaneously violated if the Lagrangian is
invariant under CP and if at the same time there is no transformation that can be
identified with a CP transformation, leaving both the Lagrangian and the
vacuum invariant. The idea of spontaneous CP violation was first proposed by 
T. D. Lee  \cite{Lee:1973iz} in the context of two Higgs doublets. In the context of the SM, 
with a single Higgs doublet, a CP transformation of the scalar doublet 
amounts to its complex conjugation and the scalar sector cannot violate CP. In models with 
several Higgs doublets  complex conjugation may be combined with a unitary transformation 
acting on the set of doublets, since this transformation leaves the kinetic energy 
term of the Lagrangian invariant. In this case the most general CP transformation is given by:
\begin{equation}
\Phi_i \overset{\mbox{CP}}{\longrightarrow} U_{ij} \Phi^\ast_j,
\end{equation}
with $U$ an arbitrary unitary matrix\footnote{Some authors refer
to this  transformation as a ``generalised" CP transformation. This is somewhat misleading since it suggests that  there is also a ``non-generalised" CP transformation.}.
This equation together with the assumption that the vacuum 
is CP invariant:
\begin{equation}
\mbox{CP} |0\rangle = |0\rangle,
\end{equation}
leads to the following condition \cite{Branco:1983tn}:
\begin{equation}
 U_{ij} \langle 0| \Phi_j |0\rangle^\ast = \langle 0| \Phi_i |0\rangle,
\label{geo}
\end{equation}
implying that there is spontaneous CP violation if none of the CP symmetries allowed by the
Lagrangian satisfy this equation. For real vevs this condition is obviously verified.
If the Lagrangian has a discrete symmetry one must take it into consideration before drawing 
conclusions. In the discussion that follows we do not take the Yukawa sector into
consideration.  We now comment  on each one of the cases presented in Table 2 concerning the
possibility of having spontaneous CP violation:

\begin{itemize}
\item
The case C-I-a is a familiar one that has been discussed long ago
in the framework of the reducible representation \cite{Branco:1983tn}. It was pointed out that
it has complex vacuum expectation values with calculable non-trivial phases, 
assuming geometrical values, entirely determined by the symmetry of the scalar potential.
These phases cannot be rotated away and yet they do not lead to spontaneous CP violation,
since there is a matrix $U$ satisfying the constraint of Eq.~(\ref{geo}), namely:
\begin{equation}
U = \left( \begin{array}{ccc}
1 & 0 & 0 \\
0 & 0 & 1 \\
0 & 1 & 0 
\end{array} \right),
\end{equation}
which is at the same time a symmetry of the potential. 

In terms of the irreducible-representation
framework we can write this solution as ($\pm i\hat w_1,\hat w_1 , 0 $ ) and 
the matrix $U$ satisfying the constraint (\ref{geo}) becomes
$U = \mbox{diag} (-1, 1, 1)$  making use of the symmetry of the potential
for $h_1 \rightarrow - h_1$.  It was shown \cite{Branco:1983tn} that 
solutions with calculable phases whose values are independent of the coupling constants of
the scalar potential do not necessarily conserve CP. Characteristic features of such solutions
in models with several Higgs doublets  as well as the interplay between symmetries 
and geometrical CP violation have been analysed by several authors \cite{deMedeirosVarzielas:2011zw,
Varzielas:2012nn,Varzielas:2012pd,Bhattacharyya:2012pi,
Ivanov:2013nla,Varzielas:2013eta,Fallbacher:2015rea}.

\item
Case C-III-a allows for a nontrivial phase which can be determined as a function of 
$\lambda_4$, and $\lambda_7$, as shown in Table~\ref{Table:Constraints-complex-I-III}. This solution violates CP spontaneously.

\item
Reasoning analogous to that for C-I-a can be applied to  cases C-III-b, C-III-d, C-III-e where again the
matrix $U$ given above, $U = \mbox{diag} (-1, 1, 1)$,  satisfies  Eq.~(\ref{geo}) 
in terms of the irreducible representation framework.  On the other hand, cases 
C-III-f and C-III-g  require $\lambda_4=0 $ and therefore the potential acquires an 
additional $SO(2)$ symmetry. In these cases $U$ can be chosen as $U = \mbox{diag} (-1, - 1, 1)$.
Case C-IV-b also requires  $\lambda_4=0 $, as a result the potential is also symmetric
under $h_2 \rightarrow - h_2$ and one can choose $U = \mbox{diag} (1, - 1, 1)$.

\item
Case C-III-c is a very interesting one. At first sight it looks as if it may violate CP 
spontaneously, however, this is not the case. In order to prove that case C-III-c does not 
violate CP spontaneously
we start from the corresponding set of vevs  $(\hat w_1 e^{i \sigma}, \hat w_2, 0)$ 
and perform a Higgs basis transformation on the 
Higgs doublets $h_1$ and $h_2$ by an $SO(2)$ rotation into:
\begin{equation}
\begin{pmatrix}
h_1^\prime \\
h_2^\prime
\end{pmatrix}   =    
\begin{pmatrix}
\cos\theta &-\sin\theta \\
\sin\theta & \cos\theta
\end{pmatrix}
\begin{pmatrix}
h_1 \\
h_2
\end{pmatrix} 
\end{equation}
such that the vevs of the new $S_3$ doublet fields now have the same modulus
and are of the form
$(ae^{i\delta_1},ae^{i\delta_2},0)$.
This requires
\begin{equation}
\tan 2 \theta = \frac{\hat w_1^2-\hat w_2^2}{2\hat w_1\hat w_2 \cos\sigma}.
\end{equation}
Obviously the Lagrangian remains invariant. Next we perform an 
overall phase rotation of the three Higgs doublets with the phase factor 
$\exp[-i(\delta_1+\delta_2)/2]$, leading now to the following vevs: 
$(ae^{i\delta},ae^{-i\delta},0)$. Making use of the symmetry for the interchange 
$h_1^\prime \leftrightarrow h_2^\prime$ we can verify Eq.~(\ref{geo})  in the following way:
\begin{equation}
\begin{pmatrix}
0 & 1 & 0 \\
1 & 0 & 0 \\
0 & 0 & 1
\end{pmatrix}
\begin{pmatrix}
ae^{i\delta} \\
ae^{-i\delta} \\
0
\end{pmatrix}^\ast
=
\begin{pmatrix}
ae^{i\delta} \\
ae^{-i\delta} \\
0
\end{pmatrix}.
\end{equation}
In terms of the initial vevs, this equation translates into
\begin{equation}
e^{i(\delta_1+\delta_2)}
\begin{pmatrix}
\cos\theta &\sin\theta & 0\\
-\sin\theta & \cos\theta & 0 \\
0 & 0 & 1
\end{pmatrix}
\begin{pmatrix}
0 & 1 & 0 \\
1 & 0 & 0 \\
0 & 0 & 1
\end{pmatrix}
\begin{pmatrix}
\cos\theta &-\sin\theta & 0 \\
\sin\theta & \cos\theta & 0 \\
0 & 0 & 1
\end{pmatrix}
\begin{pmatrix}
\hat w_1e^{i\sigma} \\
\hat w_2 \\
0
\end{pmatrix}^\ast
=
\begin{pmatrix}
\hat w_1e^{i\sigma} \\
\hat w_2 \\
0
\end{pmatrix},
\end{equation}
or
\begin{equation}
e^{i(\delta_1+\delta_2)}
\begin{pmatrix}
\sin2\theta &\cos2\theta & 0\\
\cos2\theta & -\sin2\theta & 0 \\
0 & 0 & 1
\end{pmatrix}
\begin{pmatrix}
\hat w_1e^{i\sigma} \\
\hat w_2 \\
0
\end{pmatrix}^\ast
=
\begin{pmatrix}
\hat w_1e^{i\sigma} \\
\hat w_2 \\
0
\end{pmatrix}.
\end{equation}
Notice that $(ae^{i\delta},ae^{-i\delta},0)$ is a special case of the PS vacuum, given in Eq.~(5.10).
We have checked that this case does not lead to spontaneous CP violation
even when  the two soft breaking terms discussed in
section~\ref{Sec:lambda4} are included.

\item
C-III-f,g are discussed above together with C-III-b, CP is not spontaneously violated.

\item
It is clear from this discussion that, in general, C-III-h can
violate CP and as we can see from Table~\ref{Table:Constraints-complex-I-III} the phase is determined
by parameters of the potential.

\item
For C-III-i we can verify Eq.~(\ref{geo}) 
in the reducible representation framework with $U$ acting as a permutation
between the two vevs with modulus $y$ and there is no spontaneous
CP violation. 

\item
Solution C-IV-a  is in fact real, as discussed in section~\ref{Sec:ComplexVacua}, since it requires
$\lambda_4$ and $\lambda_7$  to be zero and therefore
CP is not violated.

\item
C-IV-b is discussed above together with C-III-b, CP is not spontaneously violated.

\item
Case C-IV-c has no SO(2) symmetry because $\lambda_4$ is different from zero. As a result 
Eq.~(\ref{geo})  cannot be verified in general and therefore CP can be violated.

\item
For C-IV-d again $\lambda_4$ and $\lambda_7$ must be zero and the same reasoning
followed for  C-IV-a leads to the conclusion that CP is not violated. 

\item
For C-IV-e the reasoning is similar to the one in case C-III-c.
In order to prove that C-IV-e does not violate CP spontaneously we start with the corresponding set 
of vevs: $(\hat w_1 e^{i\sigma_1},\hat w_2e^{i\sigma_2},\hat w_S)$ where 
\begin{equation} \label{Eq:constraint}
\hat w_1=\sqrt{-\frac{\sin2\sigma_2}{\sin2\sigma_1}}\hat w_2
\end{equation}
in this phase convention. In general one should write $\sin(2\sigma_1-2\sigma_S)$ and $\sin(2\sigma_2-2\sigma_S)$ in the latter relation, where $\sigma_S$ would be the phase of the 
third vev.
We now perform an SO(2) rotation, similar to the one specified above, with
\begin{equation}
\tan 2 \theta = \frac{\hat w_1^2-\hat w_2^2}{2\hat w_1\hat w_2 \cos(\sigma_1-\sigma_2)},
\end{equation}
which once again will lead to equal moduli for the $S_3$ doublet fields. In this case, the 
vevs will acquire the form
$(be^{i\gamma_1},be^{i\gamma_2},\hat w_S)$. Unlike in case C-III-c, an overall phase rotation 
would also affect the vev of $h_S$. However, it turns out that condition 
(\ref{Eq:constraint}) enforces
\begin{equation}
\gamma_1+\gamma_2=0.
\end{equation}
This SO(2) rotation takes us to the PS vacuum.

\item
For C-IV-f there is no symmetry of the potential allowing to verify Eq.~(\ref{geo})  and
therefore CP can be  violated.

\item
Solution C-V looks like the most general case but the constraints imposed
on the parameters of the potential make it equivalent 
to a real set of vacua as discussed in section~\ref{Sec:ComplexVacua}, so that 
there is no spontaneous CP violation.
\end{itemize}

The PS vacuum specified by Eq.~(\ref{PS}) requires $\lambda_4 =0 $ and therefore 
there is symmetry under the interchange of the components of the $S_3$ doublet. As a result, 
it is possible to verify Eq.~(\ref{geo})  and CP is conserved.

The IN vacuum specified by  Eq.~(\ref{IN}) also requires $\lambda_4 =0 $.  It is clear from the
previous discussion that this solution does not lead to spontaneous CP violation. In fact, 
from Eqs.~(\ref{INa})  and (\ref{INb}) it is clear that  the allowed region of parameter space 
where this solution minimises the potential is such that either the phase $\sigma$ can be rotated 
away and therefore is not physical (IN-b) or it is fixed as $\pm \pi /2$ (IN-a) falling into  one
of the cases  C-III-f or C-III-g which were already discussed above.

We summarise these cases in Table~\ref{Table:CPV}. In this table, we also indicate whether or 
not $\lambda_4$ is equal to zero. One may conclude that $S_3$ symmetric models with 
$\lambda_4 =0$ cannot violate CP spontaneously. Still, there are 
cases with $\lambda_4 \neq 0$ where spontaneous CP violation may 
occur in three Higgs doublet models with $S_3$ symmetry.

\begin{table}[htb]
\caption{Spontaneous CP violation}
\label{Table:CPV}
\begin{center}
\begin{tabular}{|c|c|c||c|c|c||c|c|c|}
\hline
Vacuum  &  $\lambda_4$ & SCPV & Vacuum  &  $\lambda_4$ &  SCPV 
& Vacuum &  $\lambda_4$ &  SCPV\\
\hline
C-I-a & X & no & C-III-f,g & 0 & no & C-IV-c & X & yes   \\
C-III-a & X & yes  & C-III-h & X & yes  & C-IV-d & 0 & no \\
C-III-b & 0 & no  & C-III-i & X & no & C-IV-e & 0 & no \\
C-III-c & 0 & no  & C-IV-a & 0 & no & C-IV-f & X & yes \\
C-III-d,e & X & no  & C-IV-b & 0 & no & C-V & 0 & no\\
\hline
\end{tabular}
\end{center}
\end{table}

\section{Dark matter}
\label{Sec:DM}
\setcounter{equation}{0}
Multi-Higgs models may provide viable  Dark Matter candidates
in the form of one or more inert scalars. This idea was first proposed
in the context of two-Higgs-doublet models. The extra doublet is odd
under an unbroken $Z_2$ symmetry and as a result the lightest
member is stable \cite{Deshpande:1977rw,Barbieri:2006dq}.
Different implementations of this idea
have been discussed in the literature in different contexts
\cite{Ma:2006fn,Ma:2006km,
Majumdar:2006nt, LopezHonorez:2006gr, Sahu:2007uh,
Gustafsson:2007pc, Lisanti:2007ec, Hambye:2007vf, Andreas:2008xy, Hambye:2009pw,
Cao:2007rm, Lundstrom:2008ai,Dolle:2009fn,Bonilla:2014xba,
Ilnicka:2015jba}.

Inert dark matter has also been studied in the context of three-Higgs-doublet models
without an $S_3$ symmetry \cite{Grzadkowski:2009bt,Grzadkowski:2010au,Keus:2014jha}.
Models with three Higgs doublets have a richer phenomenology than those with only two.
A strong motivation for such an  extension is  the possibility of  having CP violation in the
scalar sector \cite{Grzadkowski:2009bt,Grzadkowski:2010au}.
In these models dark matter is also stabilised via a $Z_2$ symmetry.

Dark matter has been proposed within $S_3$-symmetric models, exploiting fields that get
a vanishing vev. Models of this kind include those where the singlet plays the r\^ole of the SM Higgs,
whereas the $S_3$ doublet provides dark matter
\cite{Fortes:2014dca,Fortes:2014dia,Machado:2016hca}.
An alternative way to embed dark matter could be to have the $S_3$ singlet as inert in a solution where it has a zero vev, such as C-III-c. For the C-III-c solution a specific example would correspond to the following
$S_3$ representation assignments for the quarks:
\begin{equation}
Q_{i\text{L}}:\ (2,1), \quad u_{i\text{R}}:\ (2,1), \quad d_{i\text{R}}:\ (2,1).
\end{equation}
In order to prevent $h_S$ from coupling to the quarks we would need an additional $Z_2$ symmetry under which $h_S\to-h_S$ and all other fields remain invariant. The form for the fermion mass matrices would be
\begin{equation}
\begin{pmatrix}
-aw_1 & aw_2 & bw_1 \\
aw_2 & aw_1 & bw_2 \\
cw_1 & cw_2 & 0
\end{pmatrix},
\end{equation}
with a different set of coefficients $(a, b, c)$ for the up and down
quark sector.
Since solution C-III-c requires $\lambda_4$ to vanish, no term in the
potential will break this $Z_2$ symmetry. The C-III-c vacuum breaks spontaneously
the $SO(2)$ symmetry obtained from having $\lambda_4= 0$, therefore, one way
to obtain a realistic scalar spectrum is to include the additional soft
breaking term given by Eq.~(\ref{Eq:V_soft}). This has significant  consequences for the solution
of the new minimisation conditions. The vacuum transforms into:
\begin{equation}
(\hat w e^{i \sigma}, \hat w, 0)
\end{equation}
with
\begin{equation}
\cos \sigma = - \frac{1}{4} \nu^2 \frac{1}{\hat w^2}  \frac{1}{\lambda_2 + \lambda_3}
\end{equation}
for a well defined region of parameter space. 
This was obtained by requiring $\hat w_S=0$ and $\lambda_4=0$, but relaxing the condition 
$\lambda_2+\lambda_3=0$.
At this stage, this should be seen as a toy model. In fact it has been known since
long ago that this implementation leads to an unrealistic
$V_\text{CKM}$ matrix, with only two-by-two mixing \cite{Yamanaka:1981pa}.
A full analysis of possible realistic implementations
generating the observed fermion masses and mixing is beyond the scope of this paper.

\section{Concluding remarks}
\setcounter{equation}{0}

The $S_3$-symmetric potential, with three doublets, is specified in terms of 10 parameters. It can accommodate 2 charged Higgs pairs and 5 neutral ones. If their masses were to be specified freely, one would need 7 parameters, leaving $10-7=3$ ``free''. On the other hand, if we consider an arbitrary vacuum, then 5 minimisation conditions have to be satisfied, determining 5 parameters. This mis-match illustrates that the spectrum can not be chosen freely, it will be constrained. Alternatively, one might pick a vacuum for which not all 5 minimisation conditions are independent. This would be the case, for example, when one vev vanishes.

Residual symmetries of the potential after spontaneous symmetry breaking play a very 
important r\^ole in constraining the possibility of having spontaneous CP violation
\cite{Derman:1979nf,Ivanov:2014doa,Branco:2015bfb}.  In 
Ref.~\cite{Derman:1979nf} it is proved that real vacua of $S_3$ symmetric 3HDM
always preserve an $S_2$ symmetry, whilst  constraints on complex-valued minima are 
much less severe and there are complex minima which totally break the $S_3$ symmetry.

The transition from real to complex vacua is not trivial. Our work is done
in the context of explicit CP conservation. Table~\ref{Table:real} illustrates a point that
had already been emphasized in Ref.~\cite{Derman:1979nf}, which is that for real vacua, in the
reducible representation framework, without imposing a condition equivalent
to $\lambda_4 =0$,  the only allowed solutions with all three vevs different from each other are $(x, -x, 0)$ and their permutations. This is the reason why real vacua always preserve an $S_2$ symmetry. Complex vacua, on the other hand, can evade this restriction as can be seen, for example, from solution C-IV-f.

In this work we focused our attention on the study of the scalar potential. The first necessary step 
to render such models realistic is to specify how the fermions
transform under $S_3$ and how to generate a realistic CKM matrix 
\cite{Canales:2013cga, Das:2015sca} (see also Ref.~\cite{Wyler:1979fe}).

\section*{Acknowledgments}
It is a pleasure to thank Gustavo Branco for discussions.
PO and MNR thank the CERN Theory Division for hospitality and partial support, where part of this work was done.
The authors also thank the University of Bergen and CFTP/IST Lisbon, where collaboration visits took place.
PO is supported by the Research Council of Norway. This work was partially supported by Funda\c c\~ ao para a Ci\^ encia e a Tecnologia (FCT, Portugal) through the projects CERN/FIS-NUC/0010/2015, CFTP-FCT Unit 777 (UID/FIS/00777/2013) which are partially funded through POCTI (FEDER), COMPETE, QREN and EU. DEC is currently supported by a postdoctoral fellowship from FCT Unit 777.
\appendix
\section{Converting between the two frameworks}
\setcounter{equation}{0}

The potentials in the reducible-representation framework, Eqs.~(\ref{Eq:pot-original}), and the irreducible-representation framework, Eqs.~(\ref{Eq:V-DasDey}), are related as follows:
\begin{equation} \label{Eq:App:mu0_sq}
\begin{pmatrix}
\mu_0^2 \\ \mu_1^2
\end{pmatrix}
=\frac{1}{2}
\begin{pmatrix}
-2 & 2 \\
-2 & -1
\end{pmatrix}
\begin{pmatrix}
\lambda \\ \gamma
\end{pmatrix},
\end{equation}

\begin{equation}
\begin{pmatrix}
\lambda_1 \\ \lambda_2 \\ \lambda_3 \\ \lambda_4 \\
\lambda_5 \\ \lambda_6 \\ \lambda_7 \\ \lambda_8
\end{pmatrix}
=\frac{1}{12}
\begin{pmatrix}
4 & 4 & 1 & 1 & -4 & 1 & -2 & 1 \\
3[0 & 0 & -1 & 1 & 0 & 1 & 0 & -1] \\
2 & -1 & 2 & 2 & -2 & -1 & 2 & -1 \\
\sqrt{2}[4 & -2 & -2 & -2 & -1 & 1 & 1 & 1] \\
2[4 & 4 & -2 & -2 & 2 & -2 & 1 & -2] \\
2[4 & -2 & 4 & -2 & 2 & 1 & -2 & -2] \\
4 & -2 & -2 & 4 & 2 & -2 & -2 & 1 \\
4[1 & 1 & 1 & 1 & 2 & 1 & 1 & 1]
\end{pmatrix}
\begin{pmatrix}
A \\ C \\ \overline C \\ D \\ E_1 \\ E_2 \\ E_3 \\ E_4
\end{pmatrix},
\end{equation}
with the inverse
\begin{equation}
\begin{pmatrix}
\lambda \\ \gamma
\end{pmatrix}
=\frac{1}{3}
\begin{pmatrix}
-1 & -2 \\
2 & -2
\end{pmatrix}
\begin{pmatrix}
\mu_0^2 \\ \mu_1^2
\end{pmatrix},
\end{equation}

\begin{equation}
\begin{pmatrix}
A \\ C \\ \overline C \\ D \\ E_1 \\ E_2 \\ E_3 \\ E_4
\end{pmatrix}
=\frac{1}{9}
\begin{pmatrix}
4 & 0 & 4 & 4\sqrt2 & 2 & 2 & 4 & 1 \\
2[4 & 0 & -2 & -2\sqrt2 & 2 & -1 & -2 & 1] \\
2[1 & -3 & 4 & -2\sqrt2 & -1 & 2 & -2 & 1] \\
2[1 & 3 & 4 & -2\sqrt2 & -1 & -1 & 4 & 1] \\
2[-4 & 0 & -4 & -\sqrt2 & 1 & 1 & 2 & 2] \\
2[2 & 6 & -4 & 2\sqrt2 & -2 & 1 & -4 & 2] \\
2[-4 & 0 & 8 & 2\sqrt2 & 1 & -2 & -4 & 2] \\
4[1 & -3 & -2 & \sqrt2 & -1 & -1 & 1 & 1]
\end{pmatrix}
\begin{pmatrix}
\lambda_1 \\ \lambda_2 \\ \lambda_3 \\ \lambda_4 \\
\lambda_5 \\ \lambda_6 \\ \lambda_7 \\ \lambda_8
\end{pmatrix}.
\end{equation}

\section{Positivity}
\setcounter{equation}{0}
Das and Dey \cite{Das:2014fea} have discussed {\it necessary} positivity conditions. Here, we discuss {\it necessary} and {\it sufficient} positivity conditions, following the approach of refs.~\cite{ElKaffas:2006nt,Grzadkowski:2009bt}. 
In the general case, these are rather involved. However, for the case of $\lambda_4=0$, they can be expressed in explicit form.

\subsection{General formulation}
We start by rewriting the Higgs SU(2) doublets as:
\begin{align}
\label{Eq:normphi_i}
h_i=||h_i||{\hat{h}}_i, \quad i=1,2,S,
\end{align}
where $||h_i||$ are the norms of the spinors, 
and ${\hat{h}}_i$ are unit spinors.
We let the norms of Eq.~(\ref{Eq:normphi_i}) be parametrised as follows:
\begin{equation}
\label{Eq:parametrization1}
||h_1||=r\cos\gamma\sin\theta, \qquad
||h_2||=r\sin\gamma\sin\theta, \qquad
||h_S||=r\cos\theta.
\end{equation}
The complex product between two different unit spinors will be a 
complex number with modulus less than or equal to unity, i.e.
\begin{equation}
\label{Eq:parametrization2}
{\hat{h}}_2^\dagger\cdot{\hat{h}}_1=\rho_3e^{i\theta_3}, \qquad
{\hat{h}}_S^\dagger\cdot{\hat{h}}_2=\rho_1e^{i\theta_1}, \qquad
{\hat{h}}_1^\dagger\cdot{\hat{h}}_S=\rho_2e^{i\theta_2}.
\end{equation}
Using this parametrisation\footnote{Note that this parametrisation is unrelated to that used in the body of the paper. In particular, $\rho_1$, $\rho_2$ and $\rho_3$ do not here refer to the vacuum expectation values.}, we can write:
\begin{align}
h_1^\dagger h_1&=r^2\cos^2\gamma\sin^2\theta, \quad 
h_2^\dagger h_2=r^2\sin^2\gamma\sin^2\theta, \quad
h_S^\dagger h_S=r^2\cos^2\theta,\nonumber \\
h_2^\dagger h_1&=r^2\cos\gamma\sin\gamma\sin^2\theta
\rho_3e^{i\theta_3}, \quad
h_1^\dagger h_2=r^2\cos\gamma\sin\gamma\sin^2\theta\rho_3e^{-i\theta_3},
\nonumber\\
h_S^\dagger h_2&=r^2\sin\gamma\sin\theta\cos\theta
\rho_1e^{i\theta_1}, \quad
h_2^\dagger h_S=r^2\sin\gamma\sin\theta\cos\theta\rho_1e^{-i\theta_1},
\nonumber\\
h_1^\dagger h_S&=r^2\cos\gamma\sin\theta\cos\theta
\rho_2e^{i\theta_2}, \quad
h_S^\dagger h_1=r^2\cos\gamma\sin\theta\cos\theta\rho_2e^{-i\theta_2},
\end{align}
where $r\geq0$, $\gamma\in[0,\pi/2]$, $\theta\in[0,\pi/2]$,
$\rho_i\in[0,1]$ and $\theta_i\in[0,2\pi\rangle$.

The potential can now be written as
\begin{equation}
V=r^4V_4+r^2V_2,
\end{equation}
with only the quartic, $V_4$, part relevant for positivity:
\begin{eqnarray}
\label{Eq:pot_4}
V_4&=&\lambda_1A_1+\lambda_2A_2+\lambda_3 A_3+\lambda_4A_4+\lambda_5A_5
+\lambda_6A_6+\lambda_7A_7+\lambda_8A_8,
\end{eqnarray}
where
\begin{eqnarray}
\label{Eq:A_i2}
A_1&=&\sin^4\theta, \\
A_2&=&-4\rho_3^2\sin^2\theta_3\sin^2\gamma\cos^2\gamma\sin^4\theta, \\
A_3&=&\left[\cos^4\gamma-2\cos^2\gamma\sin^2\gamma+\sin^4\gamma
+2\rho_3^2(1+\cos 2\theta_3)\cos^2\gamma\sin^2\gamma\right]\sin^4\theta, \\
A_4&=&2\left[2\rho_2\rho_3\cos\theta_2\cos\theta_3\cos^2\gamma+\rho_1\cos\theta_1(\cos^2\gamma-\sin^2\gamma)\right]\sin\gamma\cos\theta\sin^3\theta, \\
A_5&=&\cos^2\theta\sin^2\theta,\\
A_6&=&(\rho_1^2\sin^2\gamma+\rho_2^2\cos^2\gamma)\cos^2\theta\sin^2\theta,\\
A_7&=&2(\rho_1^2\cos 2\theta_1\sin^2\gamma+\rho_2^2\cos 2\theta_2 \cos^2\gamma)\cos^2\theta\sin^2\theta,\\
A_8&=&\cos^4\theta.
\end{eqnarray}
The positivity condition is then
\begin{equation}
V_4\geq0, \quad
\text{for all }\theta,\gamma,\rho_1,\rho_2,\rho_3,\theta_1,\theta_2,\theta_3.
\end{equation}
An alternative formulation of the positivity conditions has been given in terms of bilinears \cite{Maniatis:2014oza}.
\subsection{The necessary conditions of Das and Dey}
In Eq.~(4) of Das and Dey \cite{Das:2014fea}, they have listed seven necessary (but not sufficient) conditions for positivity. They can be reproduced by looking at the boundaries in $\theta$-$\gamma$ space.

\subsubsection{$\theta=0$}
\begin{equation}
V_4(\theta=0)>0 \Rightarrow\lambda_8>0,
\end{equation}
which is Eq.~(4b) of Das and Dey.

\subsubsection{$\gamma=0$}
\begin{equation}
V_4(\gamma=0)>0\Rightarrow(\lambda_1+\lambda_3)\sin^4\theta+\left[\lambda_5+\rho_2^2(\lambda_6+2\lambda_7\cos 2\theta_2)\right]\sin^2\theta\cos^2\theta+\lambda_8\cos^4\theta>0.\nonumber
\end{equation}
Minimising this with respect to $\theta_2$ we get
\begin{eqnarray}
(\lambda_1+\lambda_3)\sin^4\theta+\left[\lambda_5+\rho_2^2(\lambda_6-2|\lambda_7|)\right]\sin^2\theta\cos^2\theta+\lambda_8\cos^4\theta>0.\nonumber
\end{eqnarray}
Minimising this with respect to $\rho_2$ we get
\begin{eqnarray}
(\lambda_1+\lambda_3)\sin^4\theta+\left[\lambda_5+\text{min}(0,\lambda_6-2|\lambda_7|)\right]\sin^2\theta\cos^2\theta+\lambda_8\cos^4\theta>0.\nonumber
\end{eqnarray}
This can be treated in the same way as was done in the 2HDM \cite{ElKaffas:2006nt}, and is equivalent to the three conditions:
\begin{equation} 
\lambda_1+\lambda_3>0,\quad \lambda_8>0,\quad \lambda_5+\text{min}(0,\lambda_6-2|\lambda_7|)>-2\sqrt{(\lambda_1+\lambda_3)\lambda_8}
\label{constraints1}.
\end{equation}
These are equivalent to Eqs.~(4b), (4c), (4e) and (4f) of Das and Dey.

\subsubsection{$\theta=\pi/2$}
\begin {align}
V_4(\theta=\pi/2)>0&\Rightarrow(\lambda_1+\lambda_3)\cos^4\gamma\nonumber\\
&+2\{(\lambda_1-\lambda_3)+\rho_3^2[(\lambda_3-\lambda_2)+(\lambda_2+\lambda_3)\cos 2\theta_3]\}\cos^2\gamma\sin^2\gamma\nonumber\\
&+(\lambda_1+\lambda_3)\sin^4\gamma>0.\nonumber
\end{align}
Minimising this with respect to $\theta_3$ we get
\begin{align}
&(\lambda_1+\lambda_3)\cos^4\gamma
+2\{(\lambda_1-\lambda_3)+\rho_3^2[(\lambda_3-\lambda_2)-|\lambda_2+\lambda_3|]\}\cos^2\gamma\sin^2\gamma\nonumber\\
&+(\lambda_1+\lambda_3)\sin^4\gamma>0,\nonumber
\end{align}
or
\begin{align}
&(\lambda_1+\lambda_3)\cos^4\gamma
+2\left[(\lambda_1-\lambda_3)+2\rho_3^2\min(-\lambda_2,\lambda_3)\right]\cos^2\gamma\sin^2\gamma\nonumber\\
&+(\lambda_1+\lambda_3)\sin^4\gamma>0.\nonumber
\end{align}
Minimising this with respect to $\rho_3$ we get
\begin {align}
&(\lambda_1+\lambda_3)\cos^4\gamma
+2\left[(\lambda_1-\lambda_3)+2\min(0,-\lambda_2,\lambda_3)\right]\cos^2\gamma\sin^2\gamma\nonumber\\
&+(\lambda_1+\lambda_3)\sin^4\gamma>0.\nonumber
\end{align}
This can be treated in the same way as in Ref.~\cite{ElKaffas:2006nt}, and is equivalent to the two conditions:
\begin{equation} 
\lambda_1+\lambda_3>0,\quad \lambda_1-\lambda_3+2\min(0,-\lambda_2,\lambda_3)>-|\lambda_1+\lambda_3|
\label{constraints2}.
\end{equation}
These are equivalent to 
\begin{equation} 
\lambda_1+\lambda_3>0,\quad \lambda_1>0,\quad \lambda_1-\lambda_2>0
\label{constraints3}.
\end{equation}
The combination of these three inequalities is equivalent to the combination of (4a), (4c) and (4d) of Das and Dey.

\subsubsection{$\gamma=\pi/2$}
\begin{align}
V_4(\gamma=\pi/2)>0\Rightarrow&(\lambda_1+\lambda_3)\sin^4\theta-2\lambda_4\rho_1\cos \theta_1 \cos\theta\sin^3\theta\nonumber\\
&+(\lambda_5+\rho_1^2(\lambda_6+2\lambda_7\cos 2\theta_1))\cos^2\theta\sin^2\theta
+\lambda_8\cos^4\theta>0. \nonumber
\end{align}
The $\lambda_4$-term of this expression complicates matters due to the factor $\cos\theta\sin^3\theta$. This has not been completely solved. We can, however, reproduce (4g) of Das and Dey by putting $\rho_1=1$, $\theta=\pi/4$ and $\theta_1=0$ and $\theta_1=\pi$, respectively.
This gives us 
\begin{equation}
\lambda_1+\lambda_3-2\lambda_4+\lambda_5+\lambda_6+2\lambda_7+\lambda_8>0\quad \text{and} \quad
\lambda_1+\lambda_3+2\lambda_4+\lambda_5+\lambda_6+2\lambda_7+\lambda_8>0, \nonumber
\end{equation}
which combine into Eq. (4g) of Das and Dey.

\subsection{Positivity for models with $\lambda_4=0$}

If we put $\lambda_4=0$, we get
\begin{eqnarray}
V_4&=&\left[(\lambda_1+\lambda_3)(\cos^4\gamma+\sin^4\gamma)\right.\nonumber\\
&&\left.\hspace*{0.5cm}+2(\lambda_1-\lambda_3+\rho_3^2(\lambda_3-\lambda_2+(\lambda_2+\lambda_3)\cos 2\theta_3))\cos^2\gamma\sin^2\gamma\right]\sin^4\theta\nonumber\\
&&+\left[\lambda_5+\rho_2^2(\lambda_6+2\lambda_7\cos 2\theta_2)\cos^2\gamma
+\rho_1^2(\lambda_6+2\lambda_7\cos 2\theta_1)\sin^2\gamma
\right]\sin^2\theta\cos^2\theta\nonumber\\
&&+\lambda_8\cos^4\theta>0.
\end{eqnarray}
We minimise this with respect to $\theta_1$, $\theta_2$ and $\theta_3$ to get
\begin{eqnarray}
V_4&=&\left[(\lambda_1+\lambda_3)(\cos^4\gamma+\sin^4\gamma)\right.\nonumber\\
&&\left.\hspace*{0.5cm}+2(\lambda_1-\lambda_3+2\rho_3^2\min(-\lambda_2,\lambda_3))\cos^2\gamma\sin^2\gamma\right]\sin^4\theta\nonumber\\
&&+\left[\lambda_5+\rho_2^2(\lambda_6-2|\lambda_7|)\cos^2\gamma
+\rho_1^2(\lambda_6-2|\lambda_7|)\sin^2\gamma
\right]\sin^2\theta\cos^2\theta\nonumber\\
&&+\lambda_8\cos^4\theta>0.
\end{eqnarray}
Next, we minimise this with respect to $\rho_1$, $\rho_2$ and $\rho_3$ to get
\begin{eqnarray} \label{Eq:V4_lambda4=0}
V_4&=&\left[(\lambda_1+\lambda_3)(\cos^4\gamma+\sin^4\gamma)\right.\nonumber\\
&&\left.\hspace*{0.5cm}+2(\lambda_1-\lambda_3+\text{min}(0,-2\lambda_2,2\lambda_3))\cos^2\gamma\sin^2\gamma\right]\sin^4\theta\nonumber\\
&&+\left[\lambda_5+\text{min}(0,\lambda_6-2|\lambda_7|)
\right]\sin^2\theta\cos^2\theta\nonumber\\
&&+\lambda_8\cos^4\theta>0.
\end{eqnarray}

First we consider the boundaries in $\gamma\theta$-plane:
\subsubsection{$\theta=0$}
Like in the previous section, this leads to
\begin{eqnarray}
\lambda_8>0.
\end{eqnarray}
\subsubsection{$\gamma=0$}
Like in the previous section, this leads to
\begin{equation} 
\lambda_1+\lambda_3>0,\quad \lambda_8>0,\quad \lambda_5+\text{min}(0,\lambda_6-2|\lambda_7|)>-2\sqrt{(\lambda_1+\lambda_3)\lambda_8}.
\end{equation}
\subsubsection{$\theta=\pi/2$}
Like in the previous section, this leads to
\begin{equation} 
\lambda_1+\lambda_3>0,\quad \lambda_1>0,\quad \lambda_1-\lambda_2>0.
\end{equation}
\subsubsection{$\gamma=\pi/2$}
\begin {equation}
(\lambda_1+\lambda_3)\sin^4\theta+(\lambda_5+\text{min}(0,\lambda_6-2|\lambda_7|))\cos^2\theta\sin^2\theta
+\lambda_8\cos^4\theta>0,\nonumber
\end{equation}
Like in the previous section, this leads to
\begin{equation} 
\lambda_1+\lambda_3>0,\quad \lambda_8>0,\quad \lambda_5+\text{min}(0,\lambda_6-2|\lambda_7|)>-2\sqrt{(\lambda_1+\lambda_3)\lambda_8}.
\end{equation}

\subsubsection{Interior points in the $\theta$-$\gamma$ plane}
Minimising (\ref{Eq:V4_lambda4=0}) with respect to $\gamma$ we find (surprisingly) that the only possibility for an interior minimum occurs when $\gamma=\pi/4$. Substituting this value of $\gamma$ into (\ref{Eq:V4_lambda4=0}) we get
\begin{eqnarray}
V_4&=&\left[\lambda_1+\text{min}(0,-\lambda_2,\lambda_3)\right]\sin^4\theta\nonumber\\
&&+\left[\lambda_5+\text{min}(0,\lambda_6-2|\lambda_7|)
\right]\sin^2\theta\cos^2\theta+\lambda_8\cos^4\theta>0.
\end{eqnarray}
Like in the previous section, this leads to
\begin{equation} 
\lambda_1+\text{min}(0,-\lambda_2,\lambda_3)>0,\quad \lambda_8>0,\quad \lambda_5+\text{min}(0,\lambda_6-2|\lambda_7|)>-2\sqrt{(\lambda_1+\text{min}(0,-\lambda_2,\lambda_3))\lambda_8}.
\end{equation}
or explicitly
\begin{eqnarray} 
&&\lambda_1>0,\quad \lambda_1-\lambda_2>0,\quad\lambda_1+\lambda_3>0,\quad \lambda_8>0,\nonumber\\
&&\lambda_5+\text{min}(0,\lambda_6-2|\lambda_7|)>-2\sqrt{\lambda_1\lambda_8},\nonumber\\
&&\lambda_5+\text{min}(0,\lambda_6-2|\lambda_7|)>-2\sqrt{(\lambda_1-\lambda_2)\lambda_8},\nonumber\\
&&\lambda_5+\text{min}(0,\lambda_6-2|\lambda_7|)>-2\sqrt{(\lambda_1+\lambda_3)\lambda_8}.
\end{eqnarray}

\section{Minimisation conditions in terms of moduli and phases}
\setcounter{equation}{0}
Here, we present explicit results for the derivatives of the potential (the minimisation conditions) in terms of moduli of the vevs, and their phases.

\subsection{Conditions in the reducible-representation framework}
In the notation of equation~(\ref{Eq:reducible-vev-notation}), 
\begin{equation}
(\rho_1,\rho_2,\rho_3)=(v_1e^{i\tau_1},v_2e^{i\tau_2},v_3e^{i\tau_3}),
\end{equation}
the derivatives with respect to moduli and phases can be written as
\begin{align}
\frac{\partial V}{\partial v_1}&= - \lambda v_1 
+ \frac{1}{2} \gamma
\left[ v_2 \cos (\tau_2-\tau_1) + v_3 \cos (\tau_1-\tau_3) \right] + A v_1^3 
+ \frac{1}{2} (C + \overline{C}) v_1 ( v_2^2 + v_3^2 )  \nonumber \\
&+ \frac{1}{2} D   v_1  
\left [ v_2^2  \cos (2 \tau_2 - 2\tau_1) + v_3^2 \cos (2 \tau_1 - 2 \tau_3) \right] \nonumber \\
&+ \frac{1}{4} E_1 
\left[ (3  v_1^2  v_2 +v_2^3)\cos (\tau_2-\tau_1) 
+(3v_1^2v_3+v_3^3)\cos (\tau_1-\tau_3) \right]  \\
 & + \frac{1}{4} ( E_2 + E_3) v_2  v_3  
 \left[ 2 v_1 \cos (\tau_3 - \tau_2) + v_2 \cos (\tau_1-\tau_3) 
 + v_3 \cos (\tau_2-\tau_1)  \right] \nonumber \\
& + \frac{1}{4} E_4 v_2  v_3 
\left[ 2  v_1 \cos (\tau_2 + \tau_3 - 2\tau_1)
+ v_2 \cos (2 \tau_2 - \tau_1 - \tau_3) 
+ v_3 \cos (\tau_1 + \tau_2 - 2 \tau_3) \right]  = 0, \nonumber \\
\frac{\partial V}{\partial \tau_1}&= \frac{1}{2} \gamma v_1
\left[ v_2  \sin (\tau_2-\tau_1) - v_3 \sin ( \tau_1 - \tau_3) \right] \nonumber \\
& + \frac{1}{2} D v_1^2
\left[ v_2^2 \sin (2 \tau_2 - 2 \tau_1) - v_3^2 \sin (2 \tau_1 - 2\tau_3) \right] \nonumber \\
&+ \frac{1}{4} E_1 v_1
\left[ (v_2^3 +v_1^2 v_2) \sin (\tau_2 - \tau_1) 
-  (v_3^3+v_3v_1^2) \sin (\tau_1 - \tau_3) \right] \label{alphaa}  \\
&-  \frac{1}{4} ( E_2 + E_3) v_1  v_2  v_3 
\left[v_2 \sin (\tau_1 - \tau_3) -  v_3 \sin(\tau_2 - \tau_1) \right] \nonumber \\
&+ \frac{1}{4} E_4 v_1  v_2  v_3 
\left[ 2 v_1 \sin (\tau_2 +  \tau_3 - 2\tau_1) 
- v_2 \sin (\tau_1 - 2 \tau_2 + \tau_3) 
- v_3  \sin (\tau_1 + \tau_2 - 2 \tau_3) \right] = 0,  \nonumber  
\end{align}
with $\partial V/\partial v_2$, $\partial V/\partial v_3$, $\partial V/\partial \tau_2$, and $\partial V/\partial \tau_3$ given by cyclic permutations of the indices 1, 2 and 3.

\subsection{Conditions in the irreducible-representation framework}
We choose $w_S$ real and parametrise the complex $w_1$ and $w_2$ in terms of moduli and phases:
\begin{equation}
(w_1,w_2,w_S)=(\hat w_1e^{i\sigma_1}, \hat w_2e^{i\sigma_2}, \hat w_S).
\end{equation}
Then the derivatives with respect to moduli can be written as
\begin{align}
\left(\frac{\partial V}{\partial \hat w_1}\right)_0
&=\mu_1^2 \hat w_1
+\lambda_1 \hat w_1 (\hat w_1^2+\hat w_2^2)
+\lambda_2 \hat w_1 \hat w_2^2[\cos(2\sigma_1-2\sigma_2)-1]
+\lambda_3 \hat w_1 [\hat w_1^2+\hat w_2^2\cos(2\sigma_1-2\sigma_2)] \nonumber \\
&+\lambda_4 \hat w_1 \hat w_2 \hat w_S[\cos(2\sigma_1-\sigma_2)+2\cos\sigma_2]
+\frac{1}{2}(\lambda_5+\lambda_6)\hat w_1 \hat w_S^2
+\lambda_7 \hat w_1 \hat w_S^2 \cos2\sigma_1, \\
\left(\frac{\partial V}{\partial \hat w_2}\right)_0
&=\mu_1^2 \hat w_2
+\lambda_1 \hat w_2 (\hat w_1^2+\hat w_2^2)
+\lambda_2 \hat w_1^2 \hat w_2[\cos(2\sigma_1-2\sigma_2)-1]
+\lambda_3 \hat w_2 [\hat w_1^2\cos(2\sigma_1-2\sigma_2)+\hat w_2^2] \nonumber \\
&+\frac{\lambda_4}{2} \hat w_S
[\hat w_1^2\cos(2\sigma_1-\sigma_2)+(2 \hat w_1^2-3\hat w_2^2)\cos\sigma_2]
+\frac{1}{2}(\lambda_5+\lambda_6)\hat w_2 \hat w_S^2
+\lambda_7 \hat w_2 \hat w_S^2 \cos2\sigma_2, \\
\left(\frac{\partial V}{\partial \hat w_S}\right)_0
&=\mu_0^2 \hat w_S
+\frac{\lambda_4}{2}\hat w_2
[\hat w_1^2\cos(2\sigma_1-\sigma_2)+(2\hat w_1^2-\hat w_2^2)\cos\sigma_2]
+\frac{1}{2}(\lambda_5+\lambda_6)(\hat w_1^2+\hat w_2^2) \hat w_S \nonumber \\
&+\lambda_7\hat w_S[\hat w_1^2\cos2\sigma_1+\hat w_2^2\cos2\sigma_2]
+\lambda_8 \hat w_S^3, \label{Eq:IRF-w_S}
\end{align}
and those with respect to angles as
\begin{align}
\left(\frac{\partial V}{\partial \sigma_1}\right)_0
&=-(\lambda_2+\lambda_3)\hat w_1^2 \hat w_2^2\sin(2\sigma_1-2\sigma_2)
-\lambda_4 \hat w_1^2 \hat w_2 \hat w_S\sin(2\sigma_1-\sigma_2)
-\lambda_7 \hat w_1^2 \hat w_S^2\sin2\sigma_1, \\
\left(\frac{\partial V}{\partial \sigma_2}\right)_0
&=(\lambda_2+\lambda_3)\hat w_1^2 \hat w_2^2\sin(2\sigma_1-2\sigma_2)
+\frac{\lambda_4}{2} \hat w_2 \hat w_S
[\hat w_1^2\sin(2\sigma_1-\sigma_2) - (2\hat w_1^2-\hat w_2^2)\sin\sigma_2] \nonumber \\
&-\lambda_7 \hat w_2^2 \hat w_S^2\sin2\sigma_2.
\end{align}

When we add the soft term discussed in section~\ref{Sec:lambda4}, these derivatives get modified as follows:
\begin{align}
\frac{\partial V}{\partial \hat w_1}&=\left(\frac{\partial V}{\partial \hat w_1}\right)_0
+\half\nu^2\hat w_2\cos(\sigma_1-\sigma_2), \\
\frac{\partial V}{\partial \hat w_2}&=\left(\frac{\partial V}{\partial \hat w_2}\right)_0
+\half\nu^2\hat w_1\cos(\sigma_1-\sigma_2), \\
\frac{\partial V}{\partial \hat w_S}&=\left(\frac{\partial V}{\partial \hat w_S}\right)_0, \\
\frac{\partial V}{\partial \sigma_1}&=\left(\frac{\partial V}{\partial \sigma_1}\right)_0
-\half\nu^2\hat w_1\hat w_2\sin(\sigma_1-\sigma_2), \\
\frac{\partial V}{\partial \sigma_2}&=\left(\frac{\partial V}{\partial \sigma_2}\right)_0
+\half\nu^2\hat w_1\hat w_2\sin(\sigma_1-\sigma_2).
\end{align}


\end{document}